\documentclass[twocolumn]{svjour3}  
\smartqed  
\def\makeheadbox{\relax}
\usepackage{graphicx}
\usepackage{mathptmx}      
\usepackage[english]{babel}
\usepackage{amsmath}
\usepackage{amsfonts}
\usepackage{amssymb}
\usepackage{natbib}
\usepackage[colorlinks, allcolors=blue, urlcolor=blue, bookmarks=false, hypertexnames=true]{hyperref}
\usepackage{subcaption} 
\usepackage{breakcites}
\usepackage{makecell}
\usepackage{microtype}
\newcommand{\squeezeup}{\vspace{-5.5mm}}

\usepackage{cite}

\journalname{
Stochastic Environmental Research and Risk Assessment}

\begin{document}
	
	\title{Area-covering postprocessing of ensemble precipitation forecasts using topographical and seasonal conditions}
	
	\author{Lea Friedli*        \and
		David Ginsbourger \and Jonas Bhend }
	
	
	\institute{Lea Friedli
		\at Institute of Earth Sciences, University of Lausanne, Switzerland  \\
		\email{lea.friedli@unil.ch} 
		\and
		David Ginsbourger \at 
		Institute of Mathematical Statistics and Actuarial Science $\&$ Oeschger Center for Climate Change Research, University of Bern, Switzerland \\
		\email{ginsbourger@stat.unibe.ch}
		\and
		Jonas Bhend \at Federal Office of Meteorology and Climatology MeteoSwiss, Zurich, Switzerland \\\email{jonas.bhend@meteoswiss.ch}}
	
	\date{}

	\maketitle
	\begin{abstract}
		Probabilistic weather forecasts from ensemble systems require statistical postprocessing to yield calibrated and sharp predictive distributions. This paper presents an area-covering postprocessing method for ensemble precipitation predictions. We rely on the ensemble model output statistics (EMOS) approach, which generates probabilistic forecasts with a parametric distribution whose parameters depend on (statistics of) the ensemble prediction. A case study with daily precipitation predictions across Switzerland highlights that postprocessing at observation locations indeed improves high-resolution ensemble forecasts, with $4.5\%$ CRPS reduction on average in the case of a lead time of $1$ day. Our main aim is to achieve such an improvement without binding the model to stations, by leveraging topographical covariates. Specifically, regression coefficients are estimated by weighting the training data in relation to the topographical similarity between their station of origin and the prediction location. In our case study, this approach is found to reproduce the performance of the local model without using local historical data for calibration. We further identify that one key difficulty is that postprocessing often degrades the performance of the ensemble forecast during summer and early autumn. To mitigate, we additionally estimate on the training set whether postprocessing at a specific location is expected to improve the prediction. If not, the direct model output is used. This extension reduces the CRPS of the topographical model by up to another $1.7 \%$ on average at the price of a slight degradation in calibration. In this case, the highest improvement is achieved for a lead time of $4$ days.

	\end{abstract}
	
	\section{Introduction}
	\label{intro}
	Today, medium-range weather forecasts are generated by Numerical Weather Prediction (NWP) systems which use mathematical (or physics-based, numerical) models of the atmosphere to predict the weather. NWP forecasts are affected by considerable systematic errors due to the imperfect representation of physical processes, limited spatio-temporal resolution, and uncertainties in the initial state of the climate system. This initial condition uncertainty and the fact that the atmosphere is a chaotic system, where small initial errors can grow into large prediction errors, make weather forecasting challenging (\citeauthor{kap1_ensemble} \citeyear{kap1_ensemble}). Therefore, attention has turned to probabilistic weather forecasting to quantify weather-dependent predictability from day to day. \\
	
	Probabilistic forecasts are generated using different forecasting scenarios (referred to as ensemble members) based on slightly perturbed initial conditions and perturbed physical parameterizations in the NWP system. Unfortunately, such ensemble forecasts are not able to capture the full forecasting uncertainty as it is difficult to represent all sources of error reliably and accurately (\citeauthor{kap2_ppneed} \citeyear{kap2_ppneed}). Hence ensemble forecasts are often biased and over-confident (\citeauthor{kap3_pp}~\citeyear{kap3_pp}). Statistical postprocessing can be used to calibrate ensemble forecasts. A proper postprocessing method providing accurate weather forecasts is fundamental for risk quantification and decision making in industry, agriculture and finance. One example is flood forecasting, where reliable precipitation forecasts are a necessary prerequisite for predicting future streamflow (e.g.\ \citeauthor{SERRA2}~\citeyear{SERRA2}).  \\
	
	The objective of statistical postprocessing is to find structure in past forecast-observation pairs to correct systematic errors in future predictions. Various approaches to postprocess ensemble predictions have been developed over the last years, a selection of them is listed for example in \citeauthor{kap3_pp}~(\citeyear{kap3_pp}). His overview covers parametric approaches that assume a predictive distribution belonging to a class of probability distributions and nonparametric approaches that avoid such distributional assumptions. For the class of parametric methods, the two main approaches he lists are \textit{Bayesian model averaging} (BMA; \citeauthor{BMA_intr} \citeyear{BMA_intr}) and \textit{Ensemble model output statistics} (EMOS; \citeauthor{EMOS_intr} \citeyear{EMOS_intr}). The BMA approach generates a predictive probability density function (PDF) using a weighted average of PDFs centred on the single ensemble member forecasts. There are numerous applications of this method, for example in the studies about ensemble precipitation postprocessing of \citet{BMA1} and \citet{BMA2}. The EMOS approach provides a predictive PDF using a parametric distribution whose parameters depend on the ensemble forecast. One of the most frequently used EMOS models is the Nonhomogeneous Gaussian regression approach (NGR; \citeauthor{EMOS_intr} \citeyear{EMOS_intr}). While in a \textit{homogeneous} regression model the variance of the predictive distribution is assumed to be constant, in the \textit{inhomogeneous} approach it is expressed as a function of the ensemble variance. The NGR model, that assumes a Gaussian predictive distribution, has been extensively applied to postprocess temperature forecasts, see for instance \citet{baran} or \citet{hemri}. For precipitation, as a non-negative quantity, EMOS with a left-censoring of the forecast distribution at zero is usually applied. A range of parametric distributions have been explored for precipitation postprocessing including the censored Generalized Extreme Value distribution (\citeauthor{EMOS1} \citeyear{EMOS1}), the censored shifted Gamma distribution (\citeauthor{EMOS2} \citeyear{EMOS2}), and the censored Logistic distribution  (\citeauthor{cNLR_intr} \citeyear{cNLR_intr}). \\
	
\begin{figure}
		\centering
		\includegraphics[width=1\linewidth]{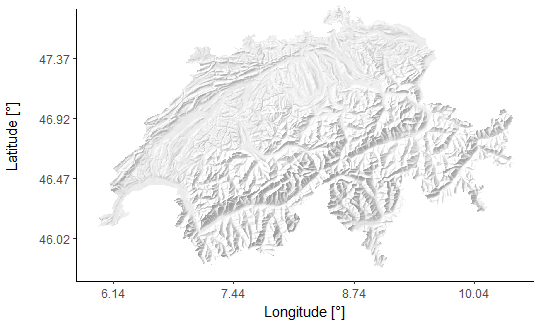}
		\caption{The relief of the study area with respect to global coordinates (WGS84)}
		\label{fig:relief}
		\squeezeup
	\end{figure}

	We seek to postprocess precipitation forecasts for all of Switzerland. With its complex topography as shown in Figure~\ref{fig:relief}, Switzerland provides a challenging case for precipitation forecasting. From a climatic perspective, the study area can be classified into different regions for which precipitation characteristics differ quite considerably. First and foremost, the Alps separate the country into a northern and southern part. The Alpine ridge often creates strong contrasts with intense precipitation on the windward slopes and dry conditions downwind. The intensity of such orography-induced precipitation also differs with much more intense precipitation frequently occuring in the south due to the advection of warm, humid air masses from the Mediterranean. The large inner-alpine valleys on the other hand are often shielded from advection of precipitation and thus tend to exhibit much drier climates than the surrounding areas. In addition to pronounced spatial variability, precipitation in Switzerland also exhibits a strong seasonal cycle. While passing fronts and large-scale precipitation dominate in the cold part of the year, in summer and autumn, convection and thunderstorms frequently occur. Convection is usually initiated away from the highest peaks on the northern and southern slope of the Alps and in the Jura mountains in the northwest. During a typical summer day, isolated showers and storms therefore start to appear there and subsequently spread according to the prevailing upper-level winds. Due to its chaotic nature and spatial heterogeneity, predicting convective precipitation is one of the key challenges in weather forecasting.\\
	
	Starting from an EMOS model, we aim to provide a postprocessing method that enables spatially comprehensive yet locally specific forecasts. To discuss alternatives to achieve this, we distinguish between \textit{global} models that use all available forecast observation pairs to estimate model coefficients, \textit{local} models that use data from a specific location only, and \textit{semi-local} models that use weighting to pool information in a suitably defined neighbourhood. The latter represents the most locally specific approach and local models therefore generally outperform global models (\citeauthor{loc_glo} \citeyear{loc_glo}). It is important to note, however, that by using local models alone, calibration at unobserved sites is not possible. Here we use a semi-local approach to enable local effects without binding the model to the stations.\\

	An ensemble postprocessing algorithm allowing for training data to be weighted individually has first been introduced by \citet{similarity2}. In their study, they calibrate ensemble precipitation forecasts using logistic regression (for a given threshold) whereby in the fitting procedure, the training data pairs are weighted with respect to the relationship of their ensemble mean and the threshold in question. Another ensemble postprocessing study where the training data is assigned with individual weights has been presented by \citet{similarity1}. Using an EMOS approach to postprocess ensemble wind speed predictions, they weight the training data pairs depending on the similarity of their location of origin and the prediction location. Thereby, they measure the similarity of two locations with distances based on the geographical location, the station climatology and the station ensemble characteristics. As an alternative to the distance based weighting approach,  \citet{similarity1} suggest to cluster the observational sites based on the same similarity measures and to perform the postprocessing for each cluster individually. The motivation behind the two semi-local approaches of \citet{similarity1} is to solve numerical stability issues of the local model, which requires long training periods since only the data of one station is used for training, they do not aim for an area-covering postprocessing method as this study does. But our study not only has another underlying motivation, we are also using new similarity measures and focus on a rich set of topographical features that are relevant for postprocessing in an area with complex topography such as Switzerland (Figure~\ref{fig:relief}).  \\
	\vspace{-0.2cm}

	Over the last years, other approaches enabling postprocessing at unobserved sites have been developed. For the interpolation technique, the postprocessing parameters of a local model are spatially interpolated using geostatistical methods. The introduction of geostatistical interpolation for a BMA postprocessing framework has been provided by \citeauthor{kleiber}~(\citeyear{kleiber}) as \textit{Geostatistical Model Averaging}. Their methodology developed for normally distributed temperature forecasts has been modified by \citet{kleiber2} to allow its application to precipitation forecasts. For EMOS models, a geostatistical interpolation procedure has been presented in \citet{scheuererb} and extended in \citet{scheuererk}. Both studies base on locally adaptive postprocessing methods avoiding location-specific parameters in the predictive distribution. Instead, they use local forecast and observation anomalies (with respect to the climatological means) as response and covariates for the regression to get a site-specific postprocessing method. Consequently, they do not have to interpolate the parameters but the anomalies. This method has been modified by \citet{dabernig} using a standardized version of the anomalies and accounting additionally for season-specific characteristics. In contrast, in the study of \citet{SERRA1}, the regression coefficients are fitted locally and then interpolated geostatistically. In their study, this two-step procedure of interpolating the coefficients is compared with an integrated area-covering postprocessing method relying on Vector Generalized Additive Models (VGAM) with spatial covariates. A comprehensive intercomparison of all the proposed approaches for area-covering postprocessing is beyond the scope of this study, instead we discuss avenues for future research in Section \ref{discussion}.  \\
	\vspace{-0.2cm}
	
	In addition to the area-covering applicability, the seasonal characteristics of the considered weather quantity present a challenge for the EMOS models. In this context, the temporal selection of the training data plays an important role. The already mentioned studies of \citet{scheuererb} and \citet{scheuererk} have been using a rolling training period of several tenths of days. This means that the model has to be refitted every day and that only part of the training data can be used for the fitting. In the study of \citet{dabernig} which is also based on anomalies, a full seasonal climatology is fitted and subtracted such that the daily fitting can be avoided and the whole training data set can be used during the regression. In the work of \citet{SERRA1}, the post-processing parameters are also fitted for every day individually. They account for seasonality by adding sine and cosine functions of seasonal covariates. We have tested similar approaches for our case study and used different choices of training periods, additional regression covariables and a weighting of the training data to account for the seasonality. Our case study highlights that a key difficulty of postprocessing ensemble precipitation forecasts lies in summer and early autumn, when in many places postprocessing leads to a degradation of the forecast quality, be it using a local or global approach. The presented seasonal approaches account for seasonal variations in the postprocessing but do not enable its renouncement. For this reason, we introduce a novel approach referred to as \textit{Pretest}. The later first evaluates whether a postprocessing at a given station in a given month is expected to improve forecast performance. A comparison of the performances shows that for our case study the Pretest approach performs best (see supplementary material for details). \\
	\vspace{-0.2cm}
	
	In summary, the aim of this paper is to provide calibrated and sharp precipitation predictions for the entire area of Switzerland by postprocessing ensemble forecasts. The postprocessing model should account for seasonal specificities and while it is developed at observation locations, it should also be applicable at unobserved locations and thereby allow to produce area-covering forecasts. The remainder of this paper is organized as follows: Section \ref{data,not,verif} introduces the data, the notation and the verification techniques. The elaboration of the postprocessing model is presented in Section \ref{ppModel}. In Section~\ref{res} we show the results of the external model validation; a discussion of the presented methodology follows in Section~\ref{discussion}. Finally, in Section~\ref{conc} the presented approach and application results are summarised in a conclusion. 
	
	\pagebreak
	\section{Data, Notation and Verification}
	\label{data,not,verif}
	
	\subsection{Data}
	\label{data}
	This study focusses on postprocessing of ensemble precipitation predictions from the NWP system \textit{COSMO-E} ({COnsortium} of Small-scale MOdelling). At the time of writing, this is the operational probabilistic high-resolution NWP system of MeteoSwiss, the Swiss national weather service. \linebreak \textit{COSMO-E} is run at 2x2km resolution for an area centered on the Alps extending from northwest of Paris to the middle of the Adriatic. An ensemble of 21 members is integrated twice daily at $00:00$ and $12:00$ UTC for five days (120 hours) into the future. Boundary conditions for the \textit{COSMO-E} forecasts are derived from the operational global ensemble prediction system of the European Centre for Medium-range Weather forecasting (ECMWF). \\
	\vspace{-0.1cm}
	
\begin{figure}[b]
		\centering
		\includegraphics[width=1\linewidth]{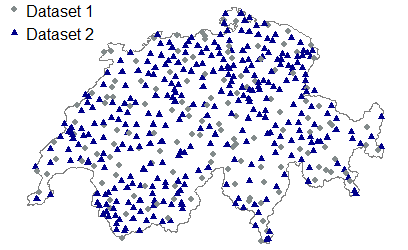}
		\caption{The $140$ stations of Dataset~1 (points) and the additional $327$ stations of Dataset~2 (triangles)}
		\label{fig:station}
	\end{figure}

	We focus on daily precipitation amounts in Switzerland. As suggested by \citet{kap11_R}, the ensemble predictions and the observations of the daily precipitation amount are square-root transformed before the postprocessing. We use observed daily precipitation at MeteoSwiss weather stations to calibrate the ensemble forecasts. This paper relies on two observation datasets, one for the elaboration of the methodology and one for the subsequent assessment of the models of choice. The datasets are presented in Table~\ref{tab:1}. The first dataset (subsequently referred to as Dataset~1) consists of ensemble forecasts and verifying observations for the daily precipitation amount between January $2016$ and July $2018$ whereby a day starts and ends at $00:00$ UTC. The data is available for $140$ automatic weather stations in Switzerland recording sub-daily precipitation. The second dataset provides the observed daily precipitation amounts of $327$ additional stations (on top of the $140$ ones of Dataset~1) that record only daily sums. For historical reasons, these daily sums are aggregated from $06:00$ UTC to $06:00$ UTC of the following day. For the purpose of uniformity, these daily limits are adopted for all stations in Dataset~2. The larger number of stations from Dataset~2 is only available from June $2016$ to July $2019$. All stations from Dataset~1 are also part of Dataset~2. The stations of both datasets are depicted in Figure~\ref{fig:station}. \\
	\vspace{-0.1cm}
	
	\begin{table*}
		\begin{center}
			\vspace{0.2cm}
			\caption{The properties of Dataset~1 and Dataset~2}
			\label{tab:1}   
			\begin{tabular}{lll}
				\hline\noalign{\smallskip}
				& \makecell[l]{\textbf{Dataset~1}} & \makecell[l]{\textbf{Dataset~2}}  \\ 
				\hline 
				\makecell[l]{Purpose} & \makecell[l]{Elaboration methodology (Section \ref{ppModel})} &  \makecell[l]{Assessment (Section \ref{res})} \\ 
				\makecell[l]{Available  months} & \makecell[l]{Jan $2016$ - Jul $2018$} & \makecell[l]{Jun $2016$ - Jul $2019$} \\ 
				\makecell[l]{Aggregation period  } & \makecell[l]{$00:00$ UTC - $00:00$ UTC} & \makecell[l]{$06:00$ UTC - $06:00$ UTC} \\ \hline
				\makecell[l]{Ensemble forecasts:} & \makecell[l]{} &  \makecell[l]{} \\
				\makecell[l]{\hspace{0.5cm} Spatial resolution} & \makecell[l]{2x2km} &  \makecell[l]{2x2km} \\
				\makecell[l]{\hspace{0.5cm} Temporal resolution} & \makecell[l]{Hourly accumulation} &  \makecell[l]{Hourly accumulation} \\ \hline
				\makecell[l]{Observations:} & \makecell[l]{} & \makecell[l]{} \\ 
				\makecell[l]{ \hspace{0.5cm} Available  stations} & \makecell[l]{$140$} & \makecell[l]{$467$ ($140$ stations of Dataset~1, $327$ additional stations)} \\ 
				\makecell[l]{\hspace{0.5cm} Temporal resolution} & \makecell[l]{Hourly accumulation} &  \makecell[l]{Hourly accumulation for the $140$ stations of Dataset~1 \\ Daily accumulation for the 327 additional stations } \\ \hline
				\makecell[l]{Model training:} & \makecell[l]{} & \makecell[l]{} \\ 
				\makecell[l]{\hspace{0.5cm} Stations} & \makecell[l]{Global models: All stations Dataset~1 (140) \\ Local model: Station of interest (1 station each)} & \makecell[l]{Global models: Stations of Datatset~1 ($140$) \\ Local model: Station of interest (1 station each) } \\ 
				\makecell[l]{\hspace{0.5cm} Months} & \makecell[l]{Cross-Validation, remove prediction month \\ ($30$ months each, using Jan $2016$ - Jul $2018$)} & \makecell[l]{$12$ months prior to prediction month \\ (using Jun $2016$ - Apr $2019$)} \\ \hline
				\makecell[l]{Model validation:} & \makecell[l]{} & \makecell[l]{} \\ 
				\makecell[l]{\hspace{0.5cm} Stations} & \makecell[l]{All stations Dataset~1 ($140$)} & \makecell[l]{Additional stations of Dataset~2 ($327$) } \\ 
				\makecell[l]{\hspace{0.5cm} Months} & \makecell[l]{Jan $2016$ - Jul $2018$} & \makecell[l]{Jun $2017$ - May $2019$} \\ 
				
				\noalign{\smallskip}\hline
			\end{tabular} 
		\end{center}
		\vspace{1cm}
	\end{table*} 
	
\begin{figure}[h]
		\begin{subfigure}[c]{\linewidth}
			\includegraphics[width=\linewidth]{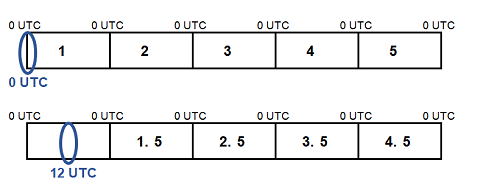}
			\subcaption{Dataset~1}
		\end{subfigure}
		\begin{subfigure}[c]{\linewidth}
			\vspace{5mm}
			\includegraphics[width=\linewidth]{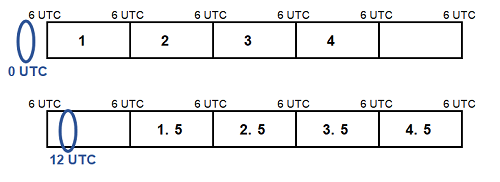}
			\subcaption{Dataset~2}
		\end{subfigure}
		\caption{The lead times of Dataset~1 and Dataset~2, which classify the forecasts with respect to the time interval between the prediction time (circle) and the predicted day}
		\label{fig:leadtimes}
	\end{figure}

	Since \textit{COSMO-E} makes forecasts for five days into the future, the different intervals between the forecast initialization time and the time for which the forecast is valid have to be taken into account. This is referred to as forecast \textit{lead time}. The possible lead times for a prediction initialized at $00:00$ UTC are $1$, $2$, $3$, $4$, $5$ days and $1.5$, $2.5$, $3.5$ and $4.5$ days for one initialized at $12:00$ UTC respectively. Figure~\ref{fig:leadtimes} illustrates the possible lead times for both datasets. For Dataset~2 the forecast lead times increase from 24 hours to 30 hours for the first day due to the different time bounds of aggregation. Also, only four complete forecasts can be derived from each \textit{COSMO-E} forecast with Dataset~2. We use Dataset~1 reduced to forecast-observation pairs with lead time equals~$3$ for the elaboration of the methodology. This selection procedure depends strongly on the dataset bearing the danger of overfitting. To assess this risk, the models of choice will be evaluated with Dataset~2. This is done for all lead times between $1$ and $4$. \\
	\vspace{-0.1cm}
	
	In addition to the forecast-observation pairs and the station location (latitude, longitude, and altitude), topographical indices derived from a digital elevation model with 25m horizontal resolution are used. The topographical indices are available on 7 grids with decreasing horizontal resolution from $25$m to $31$km describing the topography from the immediate neighbourhood (at $25$m resolution) to large-scale conditions (at $31$km). The topographical indices include the height above sea level (DEM), a variable denoting if the site is rather in a sink or on a hill (TPI), variables describing aspect and slope of the site and variables describing the derivative of the site in different directions.
	
	\subsection{Notation}
	\label{not}
	In this paper, the observed precipitation amount (at any specified location and time) is denoted as $y$. $y$ is seen as a realization of a non-negative valued random variable $Y$. The $K$ ensemble members are summarized as $\boldsymbol{x} = (x_1,...,x_K)$. Predictive distributions for $Y$ are denoted by $F$ and stand either for cumulative distribution functions (CDFs) or probability density functions (PDFs). In literature, a capital $F$ is used to refer indistinguishably to the probability measure or its associated CDF, a loose convention that we follow for simplicity within this paper. A forecast-observation pair written as $(\boldsymbol{x}_i, y_i)$ refers to a raw ensemble forecast and the corresponding observation. A pair written as $(F_i, y_i)$ generally indicates here, on the other hand, that the forecast is a postprocessed ensemble prediction. 
	
	\subsection{Verification}
	\label{verif}
	To assess a postprocessing model, the conformity of its forecasts and the associated observations is rated. In the case of probabilistic forecasts, a predictive distribution has to be compared with single observation points. We follow \citet{calibration_sharpness} by aiming for a predictive distribution maximizing the \textit{sharpness} subject to \textit{calibration}. 
	
	\paragraph{Calibration} 
	refers to the statistical consistency between forecasts and observations (\citeauthor{calibration_sharpness} \citeyear{calibration_sharpness}, \citeauthor{kap6_verif} \citeyear{kap6_verif}) and while several notions of calibration do exist (see \citeauthor{calibration_sharpness} \citeyear{calibration_sharpness} for a detailed discussion with examples highlighting their differences), the notion of \textit{probabilistic calibration} can probably be considered as the most common one. As recalled in \citet{calibration_sharpness}, "probabilistic calibration is essentially equivalent to the uniformity of the PIT values" (\textit{Probability Integral Transform}; \citeauthor{calibration}~\citeyear{calibration}). In practice, the $n$ available forecast-observation pairs $(F_i, y_i)$ with $i=1,2,...,n$ out of the test dataset are examined by having a look at the histogram of the PIT values
	\begin{equation}
	F_1(y_1),...,F_n(y_n).
	\end{equation}   
	While a flat histogram with equally populated bins is necessary for a forecasting method to be ideal, "checks for the uniformity of the PIT values have been supplemented by tests for independence" (\citeauthor{calibration_sharpness} \citeyear{calibration_sharpness}, referring to \citeauthor{fruhwirthschnatter} \citeyear{fruhwirthschnatter} and  \citeauthor{diebold}  \citeyear{diebold}). To investigate the calibration of the raw ensemble, the discrete equivalent of the PIT histogram called \textit{verification rank histogram} is used (\citeauthor{hamill_colucci} \citeyear{hamill_colucci}). It is generated by ranking the values 
	\begin{equation}
	\{ x_1, x_2,..., x_K, y \}
	\end{equation}
	of every ensemble-observation pair. A histogram of the ranks of the observations shows how they are distributed within the ensemble. Again, a flat histogram indicates a calibrated forecasting method. 
	
	\pagebreak
	\citet{hamill} pointed out that the flatness of the PIT histogram is necessary but not sufficient for a forecast to be ideal. \citet{calibration_sharpness} took these results as a motivation to aim for a predictor which maximizes the sharpness while being calibrated. Sharpness relates to the concentration of the predictive distribution; a more concentrated predictive distribution means a sharper forecast. Being a characteristic of the forecast only and not comparing it with the actual observations, sharpness is typically considered in conjunction with calibration rather than individually (\citeauthor{skill} \citeyear{skill}).
	
	\paragraph{Accuracy}
	The \textit{accuracy} of a forecast is assessed with summary measures addressing both calibration and sharpness simultaneously. These functions called \textit{Scoring Rules} map each forecast-observation pair $(F, y)$ to a numerical penalty, where a smaller penalty indicates a better forecast and vice versa (\citeauthor{kap6_verif} \citeyear{kap6_verif}). Let 
	\begin{equation}
	S : \mathcal{F} \times \mathcal{Y} \rightarrow \mathbb{R} \cup \{\infty\}
	\end{equation}
	be a Scoring Rule, where $\mathcal{Y}$ is a set with possible values of the quantity to be predicted and $\mathcal{F}$ a convex class of probability distributions on $\mathcal{Y}$. The Scoring Rule is said to be \textit{proper} relative to the class $\mathcal{F}$ if
	\begin{equation}
	\mathbb{E}_{Y \sim G}S(G,Y) \leq \mathbb{E}_{Y \sim G}S(F,Y),
	\end{equation}
	where $F, G \in \mathcal{F}$ are probability distributions and $G$ in particular is the true distribution of the random observation $Y$. The subscript $Y \sim G$ at the expected value denotes that the expected value is computed under the assumption that $Y$ has distribution $G$ (\citeauthor{calibration_sharpness} \citeyear{calibration_sharpness}). \\
	
	We will focus on two popular Scoring Rules:  The \linebreak \textit{Continuous Ranked Probability Score} (CRPS; \citeauthor{CRPS_intr} \citeyear{CRPS_intr}) and the \textit{Brier Score} (\citeauthor{Brier} \citeyear{Brier}). The CRPS is a generalization of the absolute error for probabilistic forecasts. It can be applied to predictive distributions with finite mean and is defined as follows (\citeauthor{kap6_verif} \citeyear{kap6_verif}):
	\begin{equation}
	\label{CRPS}
	CRPS(F,y) = \int_{-\infty}^{\infty} \left (F(x) - \mathbb{I}_{[y,\infty)}(x) \right)^2  \enskip dx,
	\end{equation}
	where $\mathbb{I}_A(x)$ denotes the indicator function for a set $A \subseteq \mathbb{R}$ which takes the value $1$ if $x \in A$ and $0$ otherwise. To compare the performance of the postprocessing models with the one of the raw ensemble, we need a version of the CRPS for a predictive distribution $F_{ens}$ given by a finite ensemble ${x_1, ..., x_K}$. We use the following definition by \citet{crps_ens}:
	\begin{equation}
	CRPS(F_{ens},y) = \frac{1}{K} \sum \limits_{k=1}^K |x_k - y| - \frac{1}{2K^2}\sum \limits_{k=1}^K \sum \limits_{l=1}^K |x_k-x_l|.
	\end{equation}
	In practice, competing forecasting models are compared by calculating the mean CRPS values of their predictions over a test dataset. The preferred method is the one with the smallest mean score. We use a \textit{Skill Score} as in \citet{skill} to measure the improvement (or deterioration) in accuracy achieved through the postprocessing of the raw ensemble:
	\begin{equation}
	Skill(F,F_{ens},y) = 1 - \frac{CRPS(F,y)}{CRPS(F_{ens},y)},
	\end{equation}
	where the $Skill(F,F_{ens},y)$ characterizes the improvement in forecast quality by postprocessing ($CRPS(F,y)$) relative to the forecast quality of the raw ensemble  ($CRPS(F_{ens},y)$). \\
	
	The definition of the CRPS given in Equation (\ref{CRPS}) corresponds to the integral over another Scoring Rule: The Brier Score assesses the ability of the forecaster to predict the probability that a given threshold $u$ is exceeded. The following definition of the Brier Score is taken from \citet{calibration_sharpness}: 
	\begin{equation}
	BS(F,y|u) = (F(u) - \mathbb{I}_{[y,\infty)}(u) )^2
	\end{equation}
	If the predictive distribution $F_{ens}$ is provided by a finite ensemble ${x_1, ..., x_K}$, we use the following expression for the Brier Score:
	\begin{equation}
	BS(F_{ens},y|u)  = \left\{ \Big( \frac{1}{K} \sum \limits_{k=1}^K  \mathbb{I}_{[x_k,\infty)}(u) \Big)   - \mathbb{I}_{[y,\infty)}(u)   \right\}^2. 
	\end{equation}
	
	\pagebreak
	\section{Postprocessing model}
	\label{ppModel}
	The aim of this chapter is to find a suitable postprocessing model by comparing the forecast quality of different approaches on the basis of Dataset~1.
	
	\subsection{Censored Logistic regression}
	In this section, we present a conventional ensemble postprocessing approach as a starting point for later extensions. We compared several EMOS models and Bayesian Model Averaging in a case study with Dataset~1 whereby a censored inhomogeneous Logistic regression (cNLR; \citeauthor{cNLR_intr} \citeyear{cNLR_intr}) model turned out to be the most promising approach. The performance of the models has been assessed by cross-validation over the months of Dataset~1. Of the $31$ months available, the months are removed one at a time from the training data set and the model is trained with the remaining $30$ months. Predictive performance is then assessed based on comparison between the observations at each left-out month from the models trained on the remaining months. The basic models are tested in two versions: For the global version, the model is trained with the data of all stations allowing the later application of the model to all stations simultaneously. The local version requires that models are trained individually for each station with the past data pairs of this station. More details on the alternative approaches to the cNLR model and the results from the comparison of approaches can be found in the supplementary material. \\
	
	The cNLR approach is a distributional regression model: We assume that the random variable $Y$ describing the observed precipitation amount follows a probability distribution whose moments depend on the ensemble forecast. To choose a suitable distribution for $Y$, we take into account that the amount of precipitation is a non-negative quantity that takes any positive real value (if it rains) or the value zero (if it does not rain). These properties are accounted for by appealing to a zero censored distribution. We assume that there is a latent random variable $Y^*$ satisfying the following condition (\citeauthor{crch} \citeyear{crch}):
	\begin{equation}
	Y =
	\begin{cases}
	Y^* & \text{for } Y^* > 0, \\
	0 & \text{for } Y^* \leq 0.
	\end{cases}
	\end{equation}
	In this way, the probability of the unobservable random variable $Y^*$ being smaller or equal than zero is equal to the probability of $Y$ being exactly zero. \\
	
	For the choice of the distribution of $Y^*$, we have compared different parametric distributions: a Logistic, Gaussian, Student, Generalized Extreme Value and a Shifted Gamma distribution. For the Logistic distribution, which has achieved the best results, $Y^* \sim \mathcal{L}(m,s)$ with location $m$ and scale $s$ has probability density function 
	\begin{equation}
	f(y; m,s) = \frac{\exp\Big (-\frac{y-m}{s} \Big )}{s\Big (1+\exp \Big (-\frac{y-m}{s} \Big ) \Big )^2}.
	\end{equation}
	The expected value and the variance of $Y^*$ are given by:
	\begin{equation}
	\mathbb{E}(Y^*) =  m, \quad Var(Y^*) = \frac{s^2\pi^2}{3}.
	\end{equation}
	The location $m$ and the scale $s$ of the distribution have to be estimated with the ensemble members. We note that in the \textit{COSMO-E} ensemble, the first member $x_1$ is obtained with the best estimate of the initial conditions whereas the other members are initialized with randomly perturbed initial conditions. The first member is thus not exchangeable with the other members and it is therefore reasonable to consider this member separately. Members $x_2, x_3,..., x_{21}$ are exchangeable, meaning that they have no distinct statistical characteristics. Within the location estimation, they can therefore be summarized by the ensemble mean without losing information (\citeauthor{kap3_pp} \citeyear{kap3_pp}). Taking this into account, we model the location $m$ and the scale $s$ of the censored Logistic distribution as follows:
	\begin{align}
	&m = \beta_0 + \beta_1 x_1 + \beta_2 \overline{x}, \\[1em]
	&log(s)  =  \gamma_0 + \gamma_1 SD(\boldsymbol{x}),
	\end{align}
	where $\overline{x}$ is the ensemble mean and $SD(\boldsymbol{x})$ is the ensemble standard deviation. The five regression coefficients are summarized as $\psi = (\beta_0, \beta_1, \beta_2, \gamma_0$, $\gamma_1)$. 
	
	\paragraph{Optimization and implementation}
	Let $(F_i, y_i)$ be $n$ forecast-observation pairs from our training dataset ($i=1,2,...,n$). The predictive distributions $F_i$ are censored Logistic with location and scale depending on the ensemble forecasts $\boldsymbol{x_i}$'s and the coefficient vector $\psi$. We use \textit{Scoring Rule estimation} (\citeauthor{EMOS_intr} \citeyear{EMOS_intr}) for the fitting of $\psi$. Therefore we select a suitable Scoring Rule and express the mean score of the training data pairs as a function of $\psi$. Then, $\psi$ is chosen such that the training score is minimal. In this study, the Scoring Rule of choice is the CRPS. For the implementation of the cNLR model we use the R-package \textit{crch} by \citet{crch}. 
	
	\subsection{Topographical extension}
	\label{topo_ext}
	A case study with Dataset~1 showed that the performance of the local cNLR model is better than the one of the global cNLR model (see supplementary material). Similar results have been presented for example by \citet{loc_glo} in a study about wind speed predictions. The local model, however, cannot be used for area-covering postprocessing. To improve the performance of the global model, it is enhanced with topographical covariates. The idea is to fit the regression coefficients only with training data from weather stations which are topographically similar to the prediction site. \\
	
	We assume that the training dataset consists of $n$ forecast-observation pairs $(\boldsymbol{F_i}, y_i)$ with $i=1,2,...,n$. The global cNLR model estimates $\psi$ by minimizing the mean CRPS value of all these training data pairs. To select only or favour the training pairs from similar locations, we use a weighted version of the mean CRPS value\footnote{Literature knows another kind of weighted CRPS value:  Threshold and quantile weighted versions of the CRPS are used when wishing to emphasize certain parts of the range of the predicted variable. The threshold weighted version of the CRPS is given by
		\begin{equation}
		CRPS_u(F,y) = \int_{-\infty}^{\infty} \left (F(x) - \mathbb{I}_{[y,\infty]}(x) \right)^2  u(x) \enskip dx,
		\end{equation}
		where $u$ is a non-negative weight function on the real line (\citeauthor{wCRPS} \citeyear{wCRPS}, consult their paper for the analogous definition of the quantile weighted version of the CRPS). } as the cost function $c$, which is minimized for the fitting of the coefficient vector $\psi$:
	\begin{equation}
	\label{wCRPS}
	c(\psi; s) = \sum_{i=1}^{n} w_i(s) CRPS(F_i^\psi, y_i).
	\end{equation}
	$CRPS(F_i^\psi, y_i)$ refers to the CRPS value of data pair $(F_i^\psi, y_i)$ where the predictive distribution $F_i^\psi$ depends on the coefficient vector $\psi$. We use $(F_i,y_i)$ as shorthand for $(F_i^\psi, y_i)$. The weight $w_i(s)$ of training data pair $(F_i,y_i)$ depends on the similarity between the location it originated and the location $s$ where we want to predict. We set $w_i(s)=1$ if training data pair~$i$ originated in one of the $L$ closest (be it with respect to the euclidean distance or to some other dissimilarity measure) stations to the prediction site $s$. For the training pairs from the remaining stations, we set $w_i(s)=0$. This ensures that the training dataset is restricted to the data from the $L$~stations which are most similar to the prediction site $s$. Consequently, the coefficient vector $\psi^*(s)$ which minimizes $c(\psi;s)$ depends on the location $s$. \\
	
\begin{figure}[b]
		\begin{subfigure}[c]{\linewidth}
			\includegraphics[width=\linewidth]{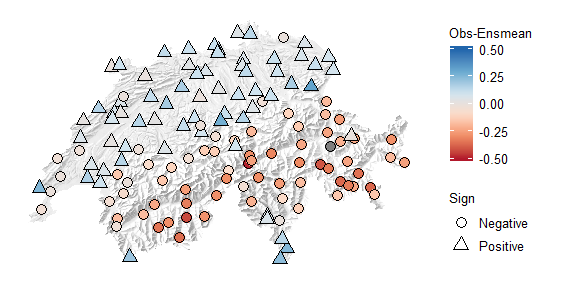}
			\subcaption{The station-wise means of the observations minus the ensemble means dependent on the coordinates of the station}
			\label{fig:coord}
		\end{subfigure}
		\begin{subfigure}[c]{\linewidth}
			\vspace{5mm}
			\includegraphics[width=\linewidth]{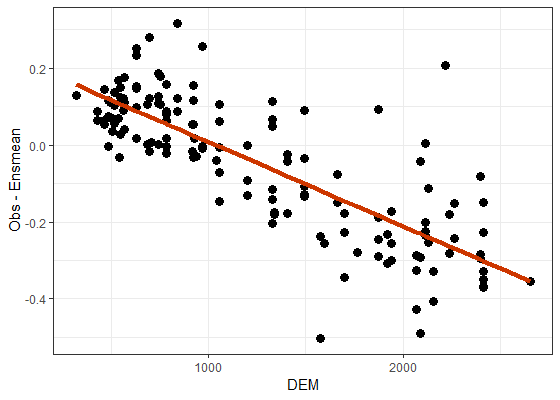}
			\subcaption{The same means as in (a) dependent on the height above the sea level (DEM) in resolution $31km$, the solid line depicts the best linear function through the points }
			\label{fig:dem}
		\end{subfigure}
		\caption{The station-wise means of the observations minus the ensemble means for the data from $2016$ of Dataset~1}
	\end{figure}

	Following \citet{similarity1}, the similarity between locations is quantified with a distance function, which, in our case, is intended to reflect the topography of the respective locations. From the topographical dataset we have about $30$ variables in $8$ resolutions at our disposal. To get an insight regarding which ones to use, we examine the topographical structure of the raw ensemble's prediction error. We compare the observed daily precipitation amounts with the ensemble means and take the station-wise averages of these differences. These preliminary analyses were made with the first year ($2016$) of Dataset~1. The mean prediction errors per station are depicted in Figure~\ref{fig:coord}. The topographical structure of the ensemble prediction error seems to be linked to the mountainous relief of Switzerland. \\
	
	In a first approach, we define the similarity of two locations via the similarity in their distances to the Alps. It turns out that such an approach depends on numerous parameters (we refer the reader to the supplementary material for more details). The proposed alternative is to focus on the variable DEM describing the height above sea level and use the values provided by a grid with low resolution (here 31km horizontal grid spacing). This ensures, that the small-scale fluctuations in the height are ignored such that the large-scale relief is represented. Figure~\ref{fig:dem} shows the same station-wise means as Figure~\ref{fig:coord}, but this time the values of each station are plotted versus their height above the sea level (DEM) in resolution $31km$. The best fit with a polynomial is achieved by modelling the ensemble mean bias as a linear function of the DEM variable; the solid line is depicting this linear function. \\
	
	The linear dependency appearing in Figure~\ref{fig:dem} motivates the following choice of a distance function to measure the similarity between two locations. Let us define a function $DEM$ which maps a location $s$ to its height above the sea level in the resolution $31$km:
	\begin{equation}
	DEM: \mathcal{D} \rightarrow \mathbb{R}, \quad s \mapsto DEM(s),
	\end{equation}
	where $\mathcal{D} \subseteq \mathbb{R}^2$ is a set with all the coordinate pairs lying in Switzerland. The similarity of locations $s_1$ and $s_2$ is then measured by the following distance:
	\begin{equation}
	\label{dDEM}
	d_{DEM}(s_1, s_2) = | DEM(s_1) - DEM(s_2) |.
	\end{equation}
	
	Based on this distance, we determine the $L$ stations of the training data set which are most similar to the prediction location, i.e. the stations which have the smallest distances $d_{DEM}$. Let
	\begin{equation}
	D_{d_{DEM}}^1(s) \leq D_{d_{DEM}}^2(s) \leq ... \leq D_{d_{DEM}}^m(s)
	\end{equation}
	be the ordered distances $d_{DEM}(s,s_j)$ between the $m$ stations from the training data and the prediction location $s$. Let $s_i$ be the location of the station where forecast-observation pair $(F_i, y_i)$ originated. Then, the weights with respect to the distance $d_{DEM}$ are defined as:
	\begin{equation}
	\label{NNwDEM}
	w_i^{L}(s) =
	\begin{cases}
	1 & \text{for } d_{DEM}(s,s_i) \leq D_{d_{DEM}}^L(s), \\
	0 & \text{otherwise}.
	\end{cases}
	\end{equation}
	These weights ensure that the data pairs, which originated at one of the $L$ most similar stations, get weight $1$ and the remaining get weight $0$. \\
	
	Besides this approach, several other topographical extensions of the cNLR model have been tested (with Dataset~1): For their spatial modelling, \citet{SERRA1} propose to vary the postprocessing parameters by expressing them as a function of spatial covariates. We have applied a similar approach and integrated the topographical covariates in the location estimation of the cNLR model. To reduce the number of predictor variables, the topographical variables have been summarized by \textit{Principal Components}. Additionally, we used the \textit{glinternet} algorithm of \citet{glinternet} to uncover important additive factors and interactions in the set of topographical variables. A more basic weighted approach has been based on Euclidean distances in the ambient (two and three dimensional) space. All extensions of the cNLR model have been compared with the local and the global fit of this very model. 
	
	As the target of this work is to develop an area-covering postprocessing method, the extended and the global models are trained with data where the predicted month and the predicted station are left out. This simulates the situation where postprocessing must be done outside a weather station, i.e. without past local measurements. The training period is set to the last year ($12$ months) before the prediction month. Consequently, forecasting performance can only be assessed with (test) data from $2017$ and $2018$. The case study with Dataset~1 showed that all other topographical approaches are less successful than the DEM approach, more details and the results can be found in the supplementary material (in the section about extension approaches). 
	
	\subsection{Seasonal extension}
\begin{figure}[t]
		\includegraphics[width=\linewidth]{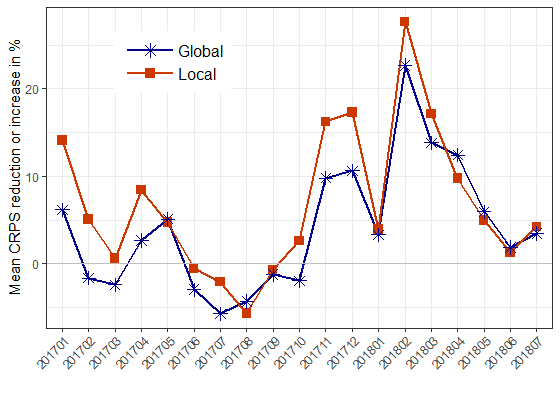}
		\caption{The monthly skill of the local and the global cNLR model which compares the monthly mean CRPS value of the model with the one of the raw ensemble, the values describe the reduction or increase of the mean CRPS value in percent}
		\label{fig:Season}
	\end{figure}

	In addition to our efforts to quantify similarities between locations, we also aim to investigate ways of further improving postprocessing outside of measurement networks by accounting for seasonal specificities. To examine the seasonal behavior of the local and the global cNLR model, we focus on their monthly mean CRPS values and compare them with the ones of the raw ensemble. Figure~\ref{fig:Season} shows the monthly skill of the global and the local cNLR model. We use the mean over all the stations from Dataset~1 and depict the months between January $2017$ and July $2018$ such that we have $12$ months of training data in each case. A positive skill of $10 \%$ means for example that the mean CRPS value of the raw ensemble is reduced by $10 \%$ through postprocessing, a negative skill indicates analogously that the mean CRPS value is higher after the postprocessing. The global model has negative skill in February and March $2017$ and between June and October $2017$. The values are especially low in July and August $2017$. The local model has a positive skill in most months, the postprocessing with this model decreases the forecasting performance only between June and September $2017$. We use these results as a first indication that the postprocessing of ensemble precipitation forecasts is particularly challenging in summer and early autumn. \\
	
\begin{figure}[b]
		\includegraphics[width=\linewidth]{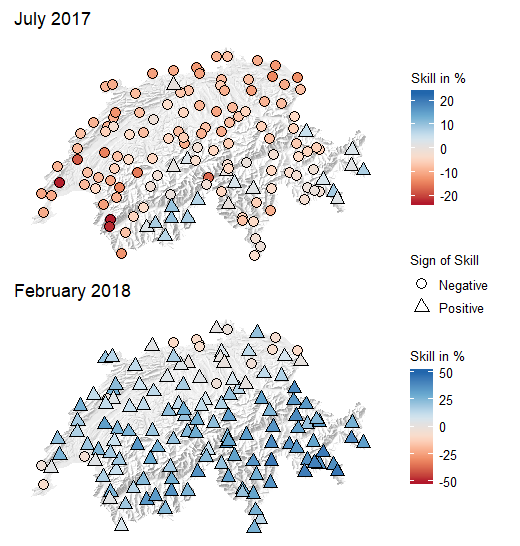}
		\caption{The station-wise skill of the global cNLR model for July $2017$ and February $2018$ which compares the mean CRPS value of the model with the one of the raw ensemble, the values describe the change in percent}
		\label{fig:Season2}
	\end{figure}

	Next, we are interested in whether there are regional differences in the model performance within a month. The global cLNR model is used as we will extend this model afterwards. We plot the maps with the station-wise means of the skill exemplary for the month with the best skill (February~$2018$) and the one with the worst (July~$2017$). The maps depicted in Figure~\ref{fig:Season2} show that the skill of the global cNLR model varies between different weather stations. Again, the structure seems to be related to the mountainous relief of Switzerland. We note that for both months the skill in the Alpine region is distinctly higher than in the flat regions. \\
	
	We use this knowledge to develop an approach which tries firstly to clarify whether the postprocessing in a given month at a given prediction location is worthwhile. The idea is to ``pretest" the model with data of similar stations and from similar months by comparing its performance with that of the raw ensemble. For this purpose, the year of training data is first reduced to the data pairs from topographically similar stations, whereby the similarity is measured with the distance $d_{DEM}$ defined in Equation (\ref{dDEM}). Afterwards, this training dataset is split into two parts: \textit{Traintrain} and \textit{Traintest}. The model is adapted a first time with the dataset Traintrain. Afterwards, the performance of this model is assessed with the second part (Traintest) by comparing the mean CRPS of the model with the mean CRPS of the raw ensemble. \\
	
	The months of the Traintest dataset are selected such that they are seasonally similar to the prediction month. To split the training dataset, three approaches are compared: 
	\begin{itemize}
		\item Pretest~1: Pretest with the same month as the prediction month from the year before (Example: January $2017$ for January $2018$)
		\item Pretest~2: Pretest with the month before the prediction month  (Example: December $2017$ for January $2018$)
		\item Pretest~3: Pretest with both of these months (Example: January $2017$ and December $2017$ for January $2018$).
	\end{itemize} 
	
	Let us define the set of indices of training data pairs out of Traintest:
	\begin{equation}
	h(i) = \{i \in \{1,2,...,n\}: (\boldsymbol{x_i}, y_i) \text{ is in Traintest}\}, 
	\end{equation}
	with cardinality~$H$. Let further $(\boldsymbol{x_i}, y_i)$ be a forecast- \linebreak observation pair where the forecast is the raw ensemble. $(F_i, y_i)$ is a pair with a postprocessed forecast. If
	\begin{equation}
	\frac{1}{H} \sum\limits_{i \in h(i)} CRPS(\boldsymbol{x_i}, y_i) \leq \frac{1}{H} \sum\limits_{i \in h(i)} CRPS(F_i, y_i),
	\end{equation}
	then the pretesting algorithm decides that the raw ensemble is not postprocessed in the given month at the given location. On the contrary if 
	\begin{equation}
	\frac{1}{H} \sum\limits_{i \in h(i)} CRPS(\boldsymbol{x_i}, y_i) > \frac{1}{H} \sum\limits_{i \in h(i)} CRPS(F_i, y_i),
	\end{equation}
	then the pretesting algorithm decides that the raw ensemble is postprocessed in the given month at the given location, the fit is done a second time with the whole year of training data.  \\
	
	The Pretest approach has been compared with several other seasonal approaches: In a basic approach, we reduce the training period to months from the same season as the prediction month. Another approach uses the sine-transformed prediction month as an additional predictor variable to model the yearly periodicity, an approach comparable to the one of \citet{SERRA1}. The third approach reduces the training data to pairs which have a similar prediction situation (quantified with the ensemble mean and the ensemble standard deviation). The methodology for the comparison has been the same as for the topographical extensions introduced in Section \ref{topo_ext}. The Pretest approach turns out to be the most promising method for our case study with Dataset~1, more details and the comparison results can be found in the supplementary material. 
	
	\pagebreak
	\subsection{Model adjustment}
	\label{modadj}
\begin{figure}[h]
		\includegraphics[width=\linewidth]{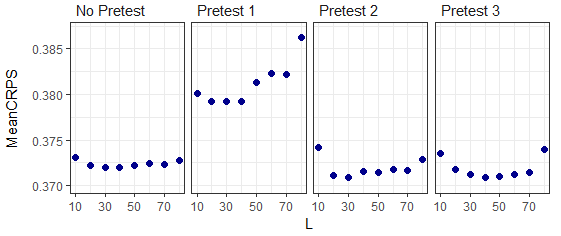}
		\caption{The mean CRPS values for the cNLR DEM (+~ Pretest) models comparing different numbers for $L$ and the different pretesting splits}
		\label{fig:L}
	\end{figure}

	For the subsequent evaluation of postprocessing models with Dataset~2, we select a few postprocessing approaches to document the impact of increasing complexity on forecast quality. We will use the raw ensemble and the local such as the global version of the cNLR model as baselines. Further on, we will evaluate the cNLR model extended by the DEM similarity (cNLR DEM). Finally, we will test this same model extended a second time with the pretest approach (cNLR DEM+PT). For the last two models, we have to fix the \linebreak amount~$L$ of similar stations we use for the topographical extension. For the last model we need to fix additionally the pretesting split. To determine the amount of similar stations in use, the numbers which are multiples of ten between $10$ and $80$ have been tested (compare Figure~\ref{fig:L}). We use the data from August $2017$ to July $2018$ (seasonally balanced) from Dataset~1 and choose the number resulting in the lowest mean CRPS. For the cNLR DEM model (no Pretest) we determine $L=40$. For the cNLR DEM+PT model, we combine the different pretesting splits with the same numbers for $L$. The cNLR DEM+PT model with the lowest mean CRPS value uses Pretest~3 and $L=40$. 
	
	\section{External validation}
	\label{res}
	This chapter presents the evaluation of the different postprocessing models. As already announced, the independent Dataset~2 is used to take into account the risk of overfitting during the elaboration of the methodology.
	
	\subsection{Methodology}
	We are interested in the area-covering performance of the models. Therefore, we are particularly interested in the performance at locations which cannot be used in model training (as no past data is available). This is the reason why we assess the models only with the $327$ additional stations of the second dataset. None of these stations have been used during the model elaboration in Section \ref{ppModel}. When determining $L$ in chapter \ref{modadj} we used a training dataset with $139$ stations ($139$ instead of $140$ as we trained without the past data from the prediction station). For this reason we carry on with using only the $140$ stations of the first dataset to train the models. This rather conservative approach could be opposed by a Cross Validation over all $467$ stations, for which, however, another choice of $L$ would probably be ideal. 
	
	The local version of the cNLR model is not able to perform area-covering postprocessing and needs the additional stations in the training from Dataset~2. Despite this, it is fitted and assessed as a benchmark here. We train the models for each of the $327$ stations and each month between June $2017$ and May $2019$. This ensures that we have one year of training data for each month and that we have seasonally balanced results. An individual fitting per station is necessary as the selection of the most similar stations used in the DEM approaches depends on the station topography. The model must also be adapted monthly, as the pretesting procedure (and the training period) depend on the prediction month. \\
	
	During the postprocessing, we used consistently the square root of the precipitation amount. The CRPS value, which is in the same unit as the observation, refers to this transformation as well. To get an idea of the actual order of magnitude, the values are converted into the original size, in which the precipitation amount is measured in $\text{mm}$. As a first step, $21$ forecasts are drawn from the fitted censored Logistic model. Afterwards, these values and the corresponding observations are squared and the mean CRPS is calculated as for the raw ensemble. The Brier Score, which assesses the ability of the forecaster to predict if a given precipitation accumulation amount is exceeded, is also evaluated for the squared sample of the predictive distribution. The thresholds used within the Brier Score  focus on three precipitation accumulations: No rain ($<$~0.1~mm/d) , moderate rain  ($>$~5~mm/d) and heavy rain  ($>$~20~mm/d). 
	
	\subsection{Results}
\begin{figure}[b]
		\centering
		\includegraphics[width=\linewidth]{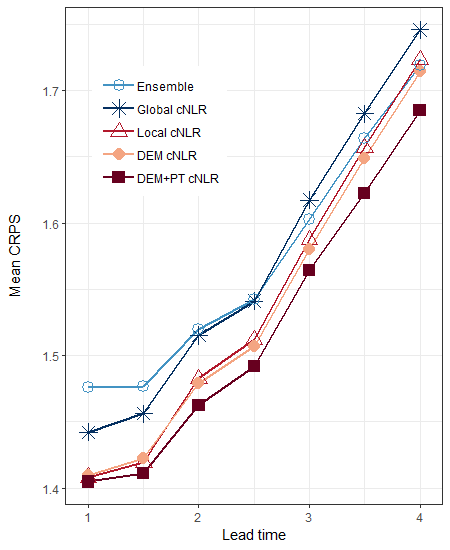}
		\caption{Mean CRPS values for the raw ensemble and the different postprocessing models dependent on the lead time, the assessment is based on the data from June $2017$ to May $2019$ of Dataset~2 }
		\label{fig:meanCRPS}
	\end{figure}

	First of all, let us give an overview of the different models. Figure~\ref{fig:meanCRPS} depicts the mean CRPS values for the different postprocessing approaches and lead times. We refer to Chapter~\ref{modadj} for a recap of the model adjustments concerning the DEM and DEM + Pretest model. For lead time $1$, the global cNLR model is able to reduce the mean CRPS value by $2.3 \%$. A further improvement is achieved by the local and the DEM model, which show equivalent performances and reduce the mean CRPS value by $4.5 \%$ compared to the raw ensemble. Even slightly better results are delivered by the DEM + Pretest model which reduces the mean CRPS value by $4.8 \%$. The skill of the global model decreases with increasing lead time. While the skill is still positive for lead times $1$, $1.5$ and $2$, the model performs roughly equally as the raw ensemble for lead time $2.5$. From lead time $3$, the mean CRPS value of the global model is even higher than the one of the raw ensemble. The local and the DEM model perform about the same for lead times between $1$ and $2.5$, for lead times above $3$ the DEM model performs slightly better. The DEM+Pretest model performs best for all lead times. It reduces the mean CRPS value between $4.8 \%$ for lead time~$1$ and $2.0 \%$ for lead time~$4$. It is noticeable that the DEM~+~Pretest model achieves a near constant improvement in the mean CRPS of approx. $0.07$ for all lead times. Relatively - i.e. as a Skill Score - this corresponds to less and less of the total forecast error. We note additionally, that the improvement which is achieved through the extension of the DEM model with the Pretest depends on the lead time. While the Pretest reduces the mean CRPS of the DEM model only by $0.4 \%$ for lead time $1$, the obtained reduction corresponds to a proportion of $1.7 \%$ for lead time~$4$. \\
	
\begin{figure*}[]
		\includegraphics[width=\linewidth]{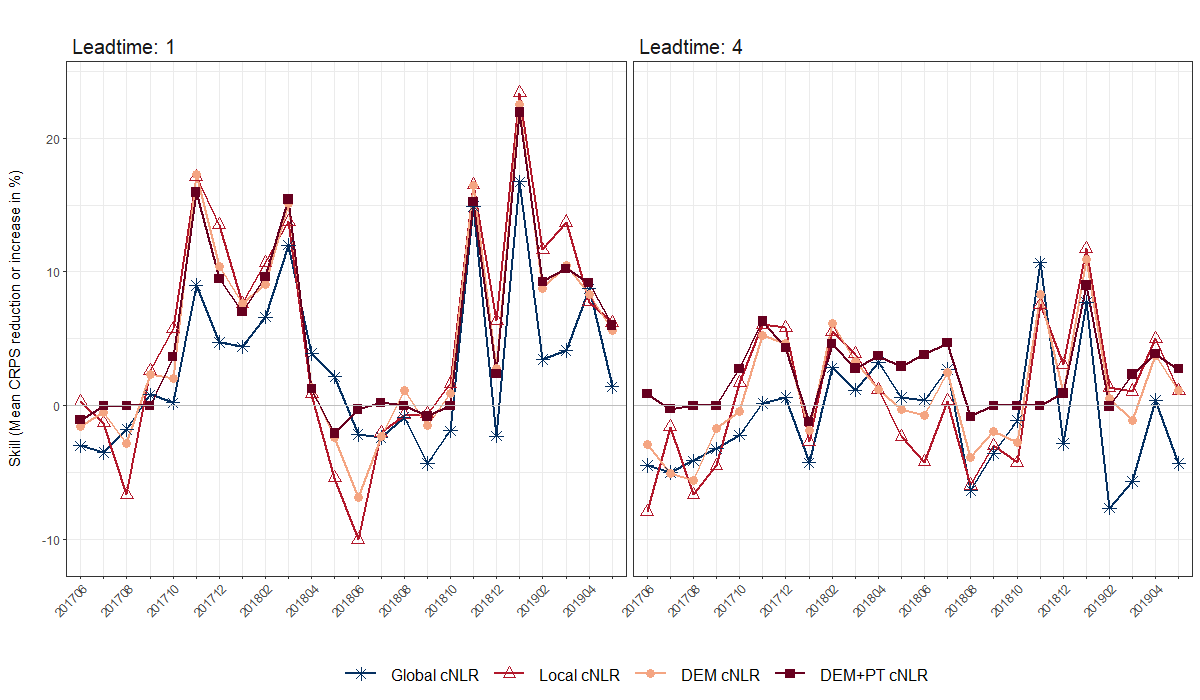}
		\caption{Reduction and increase (in $\%$) of the monthly mean CRPS value of the raw ensemble by the different postprocessing approaches, the assessment is done with the data from June $2017$ to May $2019$ of Dataset 2 }
		\label{fig:monthCRPS}
	\end{figure*}

	Figure~\ref{fig:meanCRPS} summarizes the average performance of the models over all months. To assess the seasonal performance of the different approaches, the monthly means of the Skill Score are plotted in Figure~\ref{fig:monthCRPS}. We use the raw ensemble forecast as reference and depict the results for lead time $1$ and lead time $4$. For lead time $1$, we note that the DEM + Pretest model is the only one with non-negative skill in almost all months, implying that this model only rarely degrades the quality of the ensemble prediction. While the model delivers in summer and early autumn equivalent results as the raw ensemble, the monthly mean CRPS value can be reduced by up to $20 \%$ in winter months. The same improvement is achieved with the models without Pretest, but they have a slightly worse overall performance as they degrade forecast performance during summer and autumn. For longer lead times (illustrated exemplarily for lead time $4$, right panel of Figure \ref{fig:monthCRPS}), postprocessing is less successful in improving forecast quality with forecasts in summer often deteriorating for all but the DEM + Pretest method. With pretesting the seasonal cycle in quality improvements is much less apparent for lead time $4$ than for lead time $1$. This is likely due to the combination of calibration + pretesting which is performed at individual stations, which guarantees (in expectation) that the quality of postprocessed forecasts is at least as good as that of the direct model output. If, on the other hand, there is considerable miscalibration of forecasts even if only at a few stations, this can be exploited. We also detect noticeable differences in the improvements which are achieved by extending the DEM model with the Pretest between lead times $1$ and $4$: The improvement for lead time $4$ is higher in most months, especially for June to August $2017$ and August to October $2018$.  \\
	
	We have also examined the spatial skill of the models. Therefore, we have compared the station-wise mean CRPS of the models with the one of the raw ensemble. The skill, which is very similar at neighbouring stations, increases in the Alps and is marginal or non-existent in the Swiss plateau. The resulting spatial distribution of the skill looks similar as the mean bias depicted on Figure~\ref{fig:coord}, the maps are therefore not shown here.  \\
	
\begin{figure}[b]
		\centering
		\includegraphics[width=\linewidth]{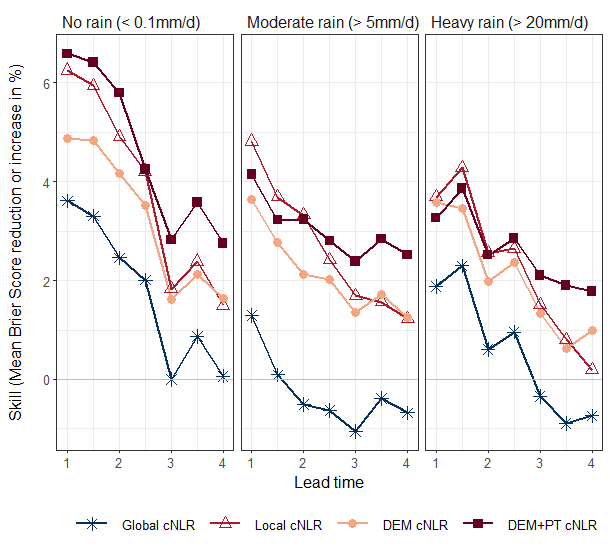}
		\caption{Skill of the different postprocessing approaches and lead times, depicted through the reduction and increase~(in~$\%$) of the mean Brier Score of the raw ensemble for the thresholds of 0.1mm/d, 5mm/d and 20mm/d. The data from June~$2017$ to May~$2019$ in use is taken from Dataset 2}
		\label{fig:Brier}
	\end{figure}

	As proposed by \citet{kap6_verif}, we use more than one Scoring Rule for the assessment of our postprocessing methods and apply the Brier Score to evaluate forecast quality of specific events. For precipitation forecasts, the ability to predict whether it will rain or not is of particular interest, this is captured by the Brier score for daily rainfall~$<$~0.1 mm (precision of observation measurements). We extend the assessment by also considering forecasts of moderate and heavy precipitation characterized by daily rainfall above 5~mm/d and 20~mm/d. Figure~\ref{fig:Brier} illustrates the skill of the different postprocessing models by comparing the mean Brier Scores of the different models, thresholds and lead times with the ones of the raw ensemble forecasts. The assessment with the Brier Score confirms that the improvement achieved with the DEM + Pretest model is higher than the one with the other models. Only for lead times 1 and 1.5 and moderate or heavy rainfall, the local model outperforms the DEM + Pretest approach. Overall, the skill decreases with increasing threshold and increasing lead time. This is to be expected given that the postprocessing focuses on improving forecasts on average and exceedances of high thresholds are relatively rare (4$\%$ of the observed daily rainfall falls in the heavy rainfall category). Also, we use square-root transformed precipitation in the optimization which further reduces the importance of heavy precipitation events in the training set. The plot confirms further that the global model performs worst, for moderate or heavy rainfall and a lead time above 1.5 respectively 2.5 even worse than the raw ensemble. As in measures of the CRPS, the local and the DEM model score comparable for all thresholds and lead times between 2.5 and 4. For smaller lead times, the local model performs better for all thresholds. \\  
	
\begin{figure}[h]
		\centering
		\includegraphics[width=\linewidth]{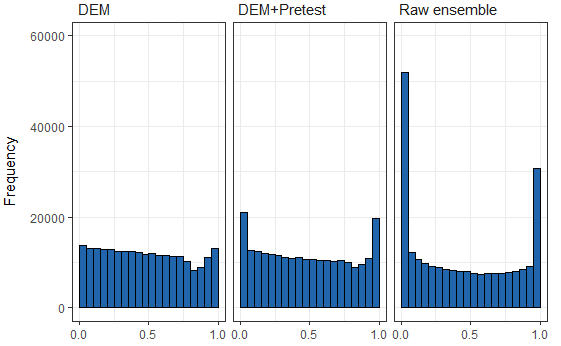}
		\caption{The PIT histograms for the DEM and DEM + Pretest models and the verification rank histogram for the raw ensemble forecasts (the lead time in use is $1$)}
		\label{fig:PIT}
	\end{figure}

	Raw ensemble forecasts are often underdispersed and have a wet bias (\citeauthor{EMOS_intr} \citeyear{EMOS_intr}). This holds for the ensemble precipitation forecasts used in this study as well. Figure~\ref{fig:PIT} (right) shows the verification rank histograms for the raw ensemble. Again, we use the data from June $2017$ to May $2019$ of Dataset~2 for this assessment and depict the results for lead time $1$. We note that the histogram for the raw ensemble has higher bins at the left and right marginal ranks and higher bins for the ranks which lie in the first half of $1,2,...21$. Therefore, it indicates that the ensemble forecasts are underdispersed and tend to have a wet bias. This raises the question whether the results of the DEM + Pretest model are still calibrated. \\
	
	To be able to evaluate the calibration of the full predictive distribution of the postprocessing models, we do not use the reverse transformation of the square root for this assessment. Additionally, we have to use a randomized version of the PIT as our predictive distribution has a discrete component (\citeauthor{kap6_verif} \citeyear{kap6_verif}):
	\begin{align}
	\lim\limits_{y \uparrow Y}F(y) + V \Big ( F(Y) - \lim\limits_{y \uparrow Y}F(y) \Big ),
	\end{align}
	where $V \sim \mathcal{U}\left([0,1]\right)$ and $y \uparrow Y$ means that $y$ approaches $Y$ from below. \newpage
	
	Figure~\ref{fig:PIT} shows the PIT histograms for the DEM and the DEM+Pretest model. As expected, the PIT histogram of the Pretest model lies between the one of the raw ensemble and the one of the DEM model without Pretest. The first and the last bins, which are higher than the other bins, indicate that the Pretest model is underdispersed. However, it seems much less gravely than for the raw ensemble. Since the remaining tested seasonal approaches produce worse results, this slight miscalibration of the DEM + Pretest model is a disadvantage we have to accept for the moment. \\
	
\begin{figure*}
		\centering
		\includegraphics[width=1\linewidth]{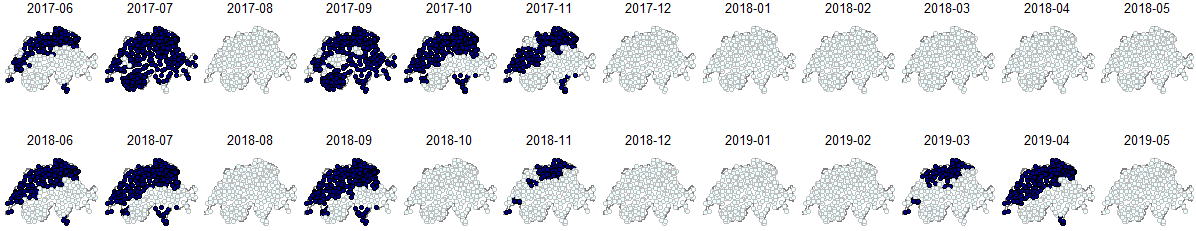}
		\caption{Maps depicting the acceptance behaviour of the DEM+Pretest model. The model evaluates for each of the $327$ stations and $24$ months of the test dataset if a postprocessing the raw ensemble seems worthwhile, the maps show when and where the model does (light point) and does not (dark point) postprocess the raw ensemble (the lead time in use is $1$) }
		\label{fig:acc}
	\end{figure*}

	Finally, we want to get an idea of the acceptance behaviour of the DEM + Pretest model. Figure~\ref{fig:acc} shows when and where the DEM + Pretest model does (light point) and does not (dark point) postprocess the raw ensemble. Again, we focus on the results of lead time $1$. The plots show that the months where the model uses the raw ensemble lie mostly in summer and autumn. The postprocessing during these months is accepted at stations which lie in bands of different widths parallel to the Alps. The model postprocesses the raw ensemble at all stations during almost all months in between December and May, the only exception are March and April $2019$. Therefore, it appears that the Pretest approach can address the seasonal difficulties of postprocessing ensemble precipitation forecasts. 
	
	\newpage
	\section{Discussion}
	\label{discussion}
	
	To enable area-covering postprocessing, we use a model that weights the training data depending on the similarity between its location of origin and the prediction location. This basic principle could be applied to any postprocessing problem where the prediction and the observation locations do not match. However, some of the choices made in this case study are quite specific and data dependent, in particular the presented procedure used to determine the number of most similar stations with which the models are trained. The models use a similarity based weighted CRPS estimator to fit the regression coefficients. The clarification of the asymptotic behaviour of such an estimator could help determining the ideal number of stations to train with and making the elaboration of the methodology less sensitive to the data in use.  \\
	
	The Pretest, which decides whether a postprocessing is worthwhile in a given setting has the disadvantage that the calibration of the resulting forecast is not guaranteed. Yet, although numerous alternative seasonal approaches have been tested, the CRPS of the Pretest model could not be levelled. In addition, making a Pretest means that the model must be adjusted twice, which is computationally expensive. But the strength of this approach is that it is fairly universally applicable also to problems outside meteorology - given that one is willing to accept some loss in model calibration for the obtained gain in accuracy.\\
	
	There are various directions in which the model could be further expanded: More meteorological information could be added such as covariates describing the large-scale flow. Further meteorological knowledge could also be incorporated by supplementing the DEM-distance with a further distinction between north and south of the Alps, for instance. The scale estimation of the final model is based only on the standard deviation of the ensemble. This estimation could be further extended with additional predictors to ensure that the ensemble dispersion is adjusted with respect to the prediction setting as well (as done for the location estimation in the alternative extension approaches). \\
	
	The evaluation with the Brier Score displays that for the case of no rain, all postprocessing models perform better than the raw ensemble. For the case of moderate rain, the global model is superseded by far by the models integrating a local aspect (local, DEM and DEM + Pretest model). The differences between the local and global models are more moderate for the case of heavy rain, but while all local approaches exceed the performance of the raw ensemble, this is not the case with the global model. Investigating further the behaviour of different approaches on the range / threshold of interest and eventually developing a postprocessing method with a focus on rare events would open exciting avenues of research. The work of \citet{extreme_events} offers an introduction to ensemble postprocessing of extreme weather events, an exemplary application for extreme rainfall intensity can be found in \citet{extreme_events2}. To avoid local fluctuations and reflect the spatial dependencies between neighbouring locations, \citet{extreme_events2} use a spatial extreme model, namely a max-stable process. The indicated potential to link postprocessing of extreme weather events to area-covering approaches is left for future research.

	\section{Conclusion} 
	\label{conc}
	The aim of this case study was to produce improved probabilistic precipitation forecasts at any place in Switzerland by postprocessing \textit{COSMO-E} ensemble forecasts enhanced with topographical and seasonal information. During the elaboration of the methodology, a censored nonhomogeneous Logistic regression model has been extended step by step; the final model combines two approaches.\\
	
	A semi-local approach is used for which only data within a neighbourhood around the prediction location are used to establish the postprocessing model. The training data used to fit the regression coefficients is weighted with respect to the similarity between its location of origin and the prediction location. This similarity is determined based on the smoothed elevation, i.e. the topographical variable DEM in a resolution of 31km. Using this approach, the weighting of the training data can be adapted for any prediction location and the model can be applied to the entire area of Switzerland thus fullfilling the first requirement of this study. \\
	
	In addition, a seasonal Pretest ensures that the model only postprocesses the raw ensemble forecast when a gain is expected - as assessed in the training sample. This extension addresses the second objective of this study and ensures that the postprocessing model accounts for seasonal specificities such as enhanced frequency of convective precipitation in the summer months. As such, the Pretest represents a flexible approach to successively integrate data-driven methods when a benchmark - here direct output from NWP - is available. This situation is expected to frequently arise in applications where training data is limited. \\
	
	The resulting final model is able to outperform a local version of the cNLR model and reduces the mean CRPS of the raw ensemble (depending on the lead time) by up to $4.8 \%$. Forecast quality might be further improved by adding meteorological and additional topographic predictors to more specifically address spatio-temporal variability of precipitation formation.
	
	\section*{Acknowledgments}
	A substantial part of the presented research was conducted when Lea Friedli was a master student at the University of Bern and David Ginsbourger was mainly affiliated with the Idiap Research Institute at Martigny. The calculations were performed on  \href{http://www.id.unibe.ch/hpc}{UBELIX}, the High Performance Computing (HPC) cluster of the University of Bern.

\newpage
\appendix
\onecolumn
\textbf{  \\}
\textbf{\huge{Supplementary material}} \\\\

\textbf{Contents}
\begin{itemize}
\item[\ref{bppm}] Basic Postprocessing models
\begin{itemize}
\item[\ref{1}] EMOS models
\item[\ref{2}] Bayesian Model Averaging
\item[\ref{3}] Results
\end{itemize}
\item[\ref{eappr}] Extension approaches
\begin{itemize}
\item[\ref{e2}] Structure ensemble prediction error
\begin{itemize}
\item[\ref{TOPOstr}] Topographical structure
\item[\ref{seasonal}] Seasonal and prediction situation dependent structure
\end{itemize}
\item[\ref{e3}] Basic extensions
\item[\ref{e4}] Results
\begin{itemize}
\item[\ref{ext1}] Topographical extensions with basic seasonality
\item[\ref{ext2}] Topographical extensions with advanced seasonality
\end{itemize}
\end{itemize}
\end{itemize}


\section{Basic Postprocessing models}
\label{bppm}
This section presents some theory and application results for several conventional ensemble postprocessing approaches. The aim is to select a suitable model as a starting point for later extensions. We compared several EMOS models and Bayesian Model Averaging in a case study with Dataset 1. The EMOS models are presented in Section~\ref{1}, the BMA model in Section~\ref{2} and the results in Section~\ref{3}. 

\subsection{EMOS models}
\label{1}
The censored nonhomogeneous regression model is applied using different distributions. In addition to the Logistic distribution presented in the main paper, a Gaussian, a Student, a Generalized Extreme Value (GEV) and a Shifted Gamma (SG) distribution have been tested out. \\

\paragraph{Censored Gaussian and censored Student distribution}
\textbf{\\}
When a censored Gaussian distribution (\textbf{cNGR} model) is used instead of a censored Logistic, the model fitting is done analogously. For a censored Student distribution (\textbf{cNSR}), which is a parametric distributions with three parameters, location and variance are estimated analogously. The additional parameter of this distribution, the number of degrees of freedom $\nu$, is estimated independently of variables from the ensemble as an intercept. The cNGR and the cNSR model are also implemented with the R-package \textit{crch} by \citet{crch} and the coefficient fit is achieved through a minimization of the average CRPS over the training dataset. \\

\paragraph{Censored GEV and censored SG distribution}
\textbf{\\}
The censored GEV and censored SG models are implemented with the R package \textit{ensembleMOS} by \citet{ensembleMOS_R}, whose functions calculate several pieces of the moment estimation internally. For this reason, most parts of the parametrization are given by the structure of the \textit{ensembleMOS} function, which is used to fit the regression models. 
For the GEV distribution, we need to estimate the location, the mean and the shape. To link the parameters of the predictive distribution to suitable predictor variables, \citet{cGEV} (and the \textit{ensembleMOS} function) add the fraction of zero members in the location, and the ensemble mean difference in the scale, respectively. They are defined as follows:
\begin{equation}
\overline{\mathbb{I}_{\{\boldsymbol{x}=0\}}} = \frac{1}{K} \sum_{k=1}^{K} \mathbb{I}_{\{x_k=0\}}, \quad 
MD(\boldsymbol{x}) = \frac{1}{K^2} \sum_{k, k'=1}^{K} |x_k - x_{k'}|.
\end{equation}
The \textit{ensembleMOS} function calculates these two quantities internally. To get reasonable results, it is therefore necessary to pass the whole ensemble to the function. To reduce the number of coefficients, the ensemble members $x_2, x_3, ..., x_{21}$ are specified as exchangeable and summarized by their mean. This leads to the following models for 
the location and scale: 
\begin{equation}
\mu = \beta_0 + \beta_1 x_1 + \beta_2 \Big( \frac{1}{20} \sum_{i=2}^{21} x_k \Big) + \beta_3 \overline{\mathbb{I}_{\{\boldsymbol{x}=0\}}},
\end{equation}
\begin{equation}
\sigma = \gamma_0 + \gamma_1 MD(\boldsymbol{x}).
\end{equation}
The shape parameter $\xi$ is estimated independently of 
the ensemble as an intercept. Again, to fit the coefficients the mean CRPS is minimized over the training dataset.   \\

For the censored SG distribution, we need to estimate the shape, the scale and the shift parameter. The shape $\kappa$ and the scale $\theta$ of the distribution are related to the location and the variance in the following way: 
\begin{equation}
\kappa = \frac{\mu^2}{\sigma^2}, \quad
\theta = \frac{\sigma^2}{\mu}.
\end{equation}
The value range of the Gamma distribution is non-negative. For this reason, the parameter $\delta$ is introduced to shift the distribution to the left and to make the censoring feasible. While this shift parameter is fitted as an intercept, the location and the scale of the distribution are linked to the ensemble in the following way (\citeauthor{cSG} \citeyear{cSG}):
\begin{equation}
\mu = \beta_0 + \beta_1 x_1 + \beta_2 \Big( \frac{1}{20} \sum_{i=2}^{21} x_k \Big),   \quad 
\sigma = \gamma_0 + \gamma_1 \overline{x}.
\end{equation}
The \textit{ensembleMOS} function used to implement the model calculates the ensemble mean $\overline{x}$ internally. In this case, a reduction of the ensemble to the control member and the ensemble mean would not lead to a nonsense estimator of the ensemble mean. Despite this, the whole ensemble is passed to the function to get comparable conditions as for the censored GEV model. Again, the coefficients are fitted by minimizing the mean CRPS over the training dataset.  \\

\paragraph{Extended Logistic regression}
\textbf{\\}
Conventional Logistic regression is dealing with binary data and can therefore not be applied in the postprocessing setting of this paper. Only if an event rather than a measured quantity is predicted, Logistic regression could be applied to postprocess an ensemble forecast. This would be the case if the event "it rains at most $q$" needs to be described. The probability of this event given that the ensemble forecast, say $\boldsymbol{X}$, equals $\boldsymbol{x}$ can be modelled as
\begin{equation}
\mathbb{P}(Y\leq q|\boldsymbol{X}=\boldsymbol{x}) = \frac{\exp(f(\boldsymbol{x}))}{1+\exp(f(\boldsymbol{x}))},
\end{equation}
where $f(\boldsymbol{x})$ is a function of chosen ensemble quantities (\citeauthor{HXLR} \citeyear{HXLR}). In what follows, this conditional probability with be denoted $\mathbb{P}(Y\leq q|\boldsymbol{x})$ for simplicity, by a slight (but common) abuse of notation. 
Using an appropriate optimization based on the training data, the coefficients of this function $f(\boldsymbol{x})$ can be determined as for the regression models presented before. The coefficients have to be fitted separately for every $q$ of interest - a large number of parameters emerges.  \\

\citet{HXLR} suggested to unify the regression functions by using one regression simultaneously for all quantiles. He adds a nondecreasing function $g(q)$ of the quantile to the function $f(\boldsymbol{x})$ and gets a single equation for every quantile:
\begin{equation}
\label{ppp}
\mathbb{P}(Y\leq q|\boldsymbol{x}) = \frac{\exp(f(\boldsymbol{x}) + g(q))}{1+\exp(f(\boldsymbol{x}) + g(q))}.
\end{equation}
The coefficients in the functions $f(\boldsymbol{x})$ and $g(q)$ have to be fitted for a selected set of observed quantiles first, but afterwards, the formula can be applied to any value of $q$. \citet{HXLR} uses $g(q)~= ~\alpha \sqrt{q}$, a choice based on empirical tests of different functions. \citet{cNLR_intr} reformulate Equation (\ref{ppp}) in the following way:
\begin{equation}
\mathbb{P}(Y \leq q|\boldsymbol{x}) = \mathbb{P}(\sqrt{Y}\leq \sqrt{q}|\boldsymbol{x}) =  \frac{\exp\Big(\frac{f'(\boldsymbol{x}) + \sqrt{q})}{1 / \alpha}\Big)}{1+\exp\Big(\frac{f'(\boldsymbol{x}) + \sqrt{q})}{1 / \alpha}\Big)},
\end{equation}
where $f'(\boldsymbol{x}) = f(\boldsymbol{x}) / \alpha$. It follows that the predictive distribution of $\sqrt{Y}$ is Logistic with location $-f'(\boldsymbol{x})$ and scale $1 / \alpha$. To add information from the ensemble spread, \citet{HXLR} replace $1 / \alpha$ by $\exp(h(\boldsymbol{x}))$ such that the variance of the Logistic distribution can be modelled through a function $h(\boldsymbol{x})$ of the ensemble as well. \\

The Extended Logistic Regression (HXLR) model can be implemented with the function \textit{hxlr} out of the R-package \textit{crch} by \citet{crch}. Again, the location of the Logistic distribution of $Y$ is assumed to depend linearly on $x_1$ and $\overline{x}$, the variance on $SD(\boldsymbol{x})$. The square root transformation of the model presented above is omitted as the whole data has been transformed in this way at the beginning. This leads to the following formula for the probability of $Y$ being smaller or equal than a quantile~$q$:
\begin{equation}
\mathbb{P}(Y \leq q|\boldsymbol{x}) = \cfrac{\exp \left( \cfrac{q - (\beta_0 + \beta_1x_1 + \beta_2 \overline{x})}{\exp(\gamma_0 + \gamma_1 SD(\boldsymbol{x}))}\right)}
{1 + \exp \left( \cfrac{q - (\beta_0 + \beta_1x_1 + \beta_2 \overline{x})}{\exp(\gamma_0 + \gamma_1 SD(\boldsymbol{x}))}\right)}.  
\end{equation}
In this paper, the coefficients $\beta_0$, $\beta_1$, $\beta_2$, $\gamma_0$ and $\gamma_1$ are fitted by minimizing the mean CRPS of the quantiles $q_{0.1}, q_{0.2},..., q_{0.9}$ from the training data. As precipitation is the quantity of interest, we have to make the model feasible by censoring the Logistic distribution after the fitting. 

\subsection{Bayesian Model Averaging}
\label{2}
Bayesian Model Averaging (BMA; \citeauthor{BMA_intr} \citeyear{BMA_intr}) generates a predictive PDF which is a weighted average of PDFs centred on the single ensemble member forecasts, which have already been bias-corrected (\citeauthor{BMA} \citeyear{BMA}). For the estimation of the predictive PDF, every ensemble member is associated with a conditional PDF $f_k(y|x_k)$ which can be imagined as the PDF of $Y$ given $x_k$ under the assumption that $x_k$ is the best prediction of the ensemble. The weighted average is defined as: 
\begin{equation}
f_{BMA}(y|\boldsymbol{x}) =   \sum \limits_{k=1}^m w_kf_k(y|x_k).
\end{equation}
\citet{BMA} specify the non-negative weight $w_k$ as the posterior probability of member $k$ being the best forecast (based on the performance of this ensemble member in the training period). The sum of the weights over all $k \in K$ is one, where $K$ is the number of ensemble members. \\
When modelling $f_k(y|x_k)$, the properties of the precipitation amount have to be considered. For this reason, \citet{BMA} apply a function consisting of two parts: One describes the probability of precipitation as a function of $x_k$, the second specifies the PDF of the amount of precipitation given that it is non-zero. \\

The implementation of the BMA model is done in the following way: Firstly, the ensemble members have to be bias-corrected with a linear regression:
\begin{equation}
x_{k}^{cor} = a_k + b_kx_{k},
\end{equation} 
where $a_k$ and $b_k$ are fitted by minimizing the average square difference between the observations and the corrected members in the training period. In the following aligns, $x_k$ is treated as the corrected version $x_k^{cor}$. \\
The conditional PDF $f_k(y|x_k)$ is modelled as:
\begin{equation}
f_k(y|x_k) 	= p_{k}\mathbb{I}(y=0) + (1-p_{k})g_{k}(y)\mathbb{I}(y > 0) .
\end{equation}
\citet{BMA} estimate the probability of no precipitation with Logistic regression:
\begin{equation}
p_{k} = \mathbb{P}(Y=0 | x_k) = \frac{\exp(a_{0,k} + a_{1,k}x_{k} + a_{2,k}I(x_{k}=0))}{1 + \exp(a_{0,k} + a_{1,k}x_{k} + a_{2,k}I(x_{k}=0))},
\end{equation}
where an indicator term for the case that the ensemble member $x_k$ is equal to zero is added. To specify the PDF $g_k(y)$ of the precipitation amount given that it is non-zero, \citet{BMA} use a Gamma distribution. The BMA model used in this paper fits the location and the scale of the Gamma distribution as follows: 
\begin{equation}
\mu_{k} = b_{0,k} + b_{1,k}x_{k}, \quad 
\sigma_{k} = c_{0} + c_{1}x_{k}.
\end{equation}
\citet{BMA} figured out that the variance parameters did not vary much between the ensemble members of their dataset (precipitation) and assumed therefore that they are constant over all members. This is adopted in this paper.  \\

Bayesian Model Averaging is implemented with the R-package \textit{ensembleBMA} by \citet{ensembleBMA_R}. The eponymous function first estimates the coefficients of $p_k$ with a Logistic regression over the training data. Then, the coefficients in $\mu_k$ are estimated by Linear Regression. At last, the coefficients in $\sigma_k$ are fitted by maximum likelihood estimation (\citeauthor{BMA} \citeyear{BMA}). To get a model which has a comparable number of coefficients as the other basic models of this paper, the ensemble is reduced to the control member $x_1$ and the ensemble mean $\overline{x}$ before applying the BMA model. 

\newpage
\subsection{Results}
\label{3}
The performance of the basic models is assessed with Dataset 1. A Cross-Validation over the months is applied. Of the $31$ months available, one month is removed from the training dataset and the model is trained with the remaining $30$ months. After that, the performance is evaluated with the month previously removed. At this point, it is still accepted that the model is trained with future data. The performance of the basic models is evaluated with the CRPS and the PIT histogram. In addition, Skill Scores are used to compare the average CRPS of the models with the one of the raw ensemble. The Skill Scores are averaged per month and per station to get a spatio-temporal impression of the model performance. In this paper, we use the square root of the predictions and the observations for the postprocessing. The assessment here is done with this transformed values such that the magnitude is not the same as in the result section of the main paper, where we reversed the transformation.\\

The basic models are tested in two versions: A global fit where all stations get the same coefficients and a local fit with separate coefficients per station. For the global version, the model is trained with the data of all stations allowing the later application of the model to all stations simultaneously. The local version requires that each station is trained individually with the past data pairs of this station. \\

\begin{figure}[h]
	\centering
	\includegraphics[width=0.8\textwidth]{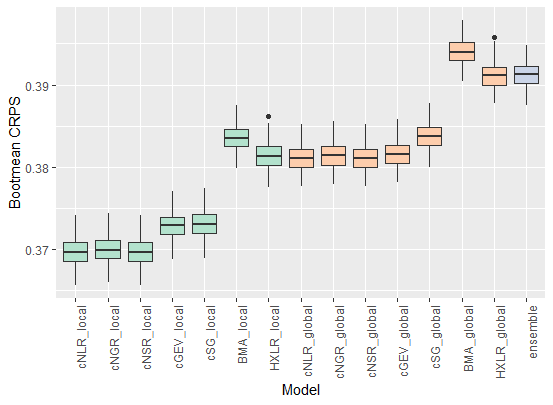} \\
	\caption{Boxplot of the bootstrapped mean CRPS values for the basic models. \newline }
	\label{fig:crps_b}
\end{figure}

First, the overall performances of the local and global versions of the basic models are compared. As proposed by \citet{kap6_verif}, a bootstrapped mean of the CRPS values is calculated per model. That means that repeatedly a sample of the same size as the test dataset is drawn (with replacement) from the CRPS values of the test dataset. In this paper, such a sample is drawn $250$ times, such that $250$ values for the mean CRPS emerge. These values are represented in the Boxplot of Figure~\ref{fig:crps_b}. The green boxes represent local model fits, the orange global ones and the blue box represents the bootstrapped mean CRPS values of the raw ensemble. It clearly appears that the local versions of the basic models perform better in terms of the mean CRPS value. The censored nonhomogeneous regression models perform best in both fits. Especially for the local version, the cNLR, cNGR and cNSR models perform better than the models using a GEV and a Shifted Gamma distribution. The BMA model has the highest mean CRPS value, which could possibly be due to the fact that the model only provides two ensemble quantities with densities. The HXLR model has the second highest mean CRPS value, it seems that it performs worse than the full version of cNLR.  \\

\newpage
To get an idea of the seasonal performance, we calculate the Skill Scores of the postprocessing models for every month~$m$. In our dataset, the months from January $2016$ to July $2018$ are available. The CRPS serves as measure of accuracy and the raw ensemble is used as reference forecast. We define $I(m) = \{i \in \{1,2,...,n\}: \text{ data pair }i \text{ originated in month } m \}$ with cardinality $M$. 
Let $(x_i, y_i)$ be a forecast-observation pair from month $m$. $x_i$ is denoting a raw ensemble forecast. $(F_i, y_i)$ is a same data pair but with a postprocessed forecast. The Skill Score of the postprocessing model for month $m$ is calculated as:
\begin{align}
\label{SSm}
1 - \frac{\frac{1}{M}\sum\limits_{i \in I(m)}CRPS(F_i, y_i)}{\frac{1}{M}\sum\limits_{i \in I(m)}CRPS(x_i, y_i)}.
\end{align}

\begin{figure}[h]
	\centering
	\includegraphics[width=0.45\textwidth]{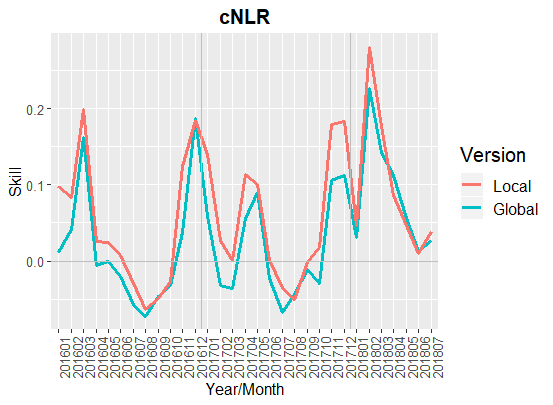} 
	\includegraphics[width=0.45\textwidth]{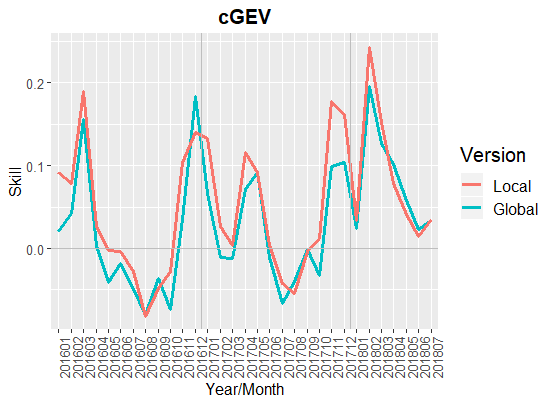} 
	\includegraphics[width=0.45\textwidth]{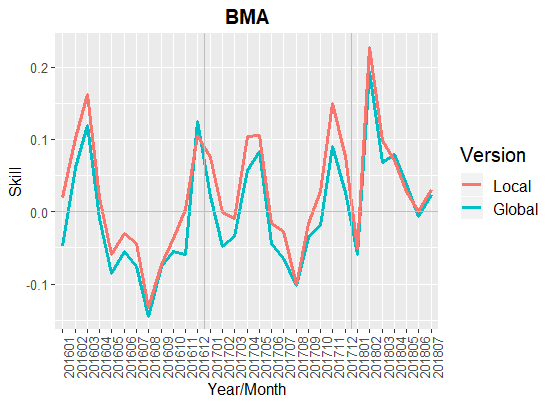} 
	\includegraphics[width=0.45\textwidth]{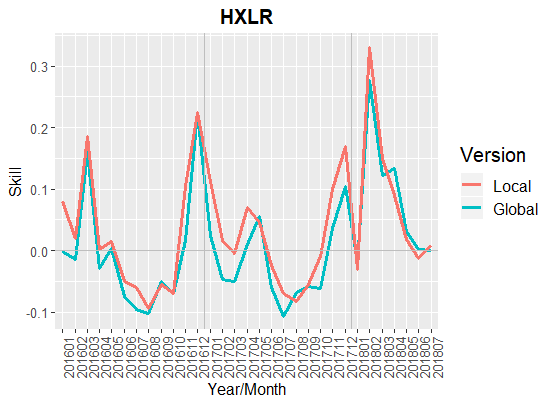} 
	\caption{Monthly Skill Score. Compares the monthly mean CRPS values of the basic postprocessing models with the ones of the raw ensemble. \newline}
	\label{fig:timeline_b}
\end{figure}

In Figure~\ref{fig:timeline_b}, the monthly skills for four of the basic models are depicted, the ones of the remaining three look similarly. We notice that all the postprocessing models have a lack of skill during summer and early autumn: The skill is negative what indicates that the raw ensemble performs better. This will be one of the difficulties to deal with in the extension of the models.   \\

\begin{figure}[h]
	\centering
	\vspace{0.5cm}
	\includegraphics[width=0.8\textwidth]{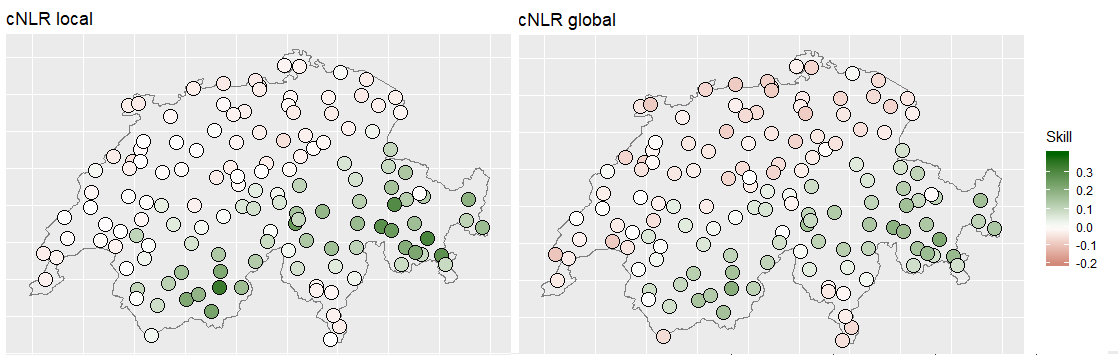} 
	\includegraphics[width=0.8\textwidth]{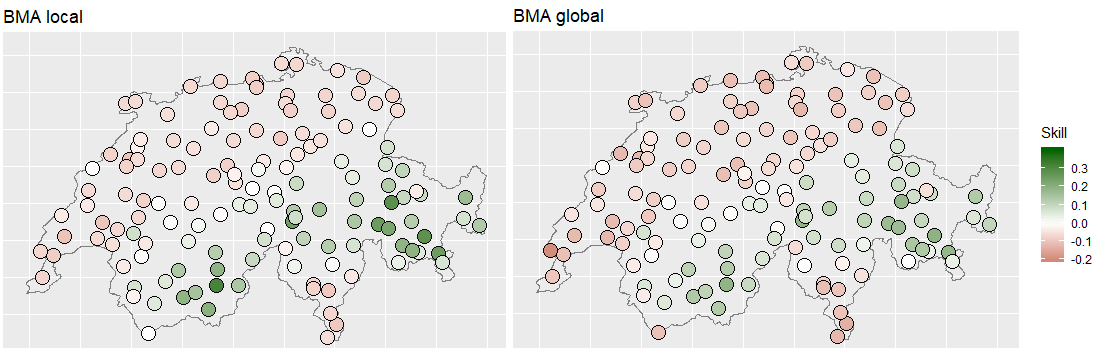} 
	\includegraphics[width=0.8\textwidth]{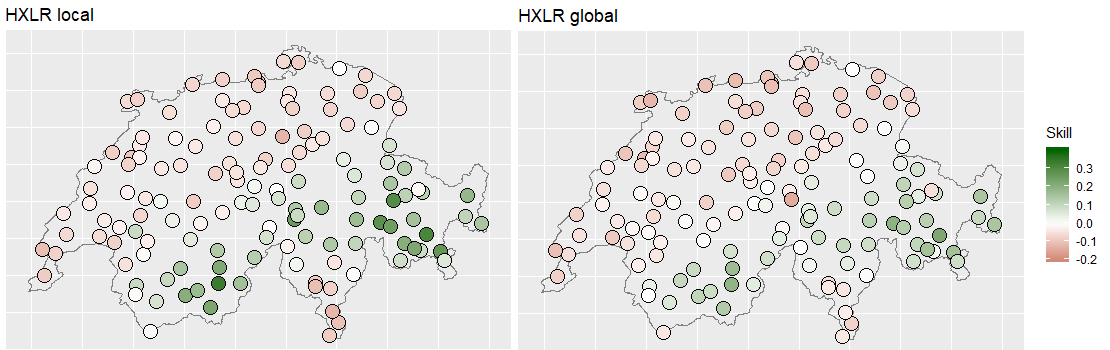} 
	\caption{Station-wise Skill Score. Compares the station-wise mean CRPS values of the basic postprocessing models with the ones of the raw ensemble. \newline}
	\label{fig:map_b}
\end{figure}

\newpage
The spatial performance of the models can be assessed by having a look at the maps of the station-wise Skill Scores. This time, the mean CRPS values of the postprocessing models respectively the raw ensemble are calculated per station. In Figure~\ref{fig:map_b}, the plots of the cNLR, the BMA and the HXLR model are printed. The maps of the other censored nonhomogeneous regression models look similarly. It is evident that the postprocessing performs better in the Alps than in the lowland. For the local version of the cNLR model, the average skill in the lowland is around zero, while the other two local models and all the global models have negative skill there. We will make use of this spatial structure in the later extensions. \\

\begin{figure}[h]
\centering
\includegraphics[width=1\textwidth]{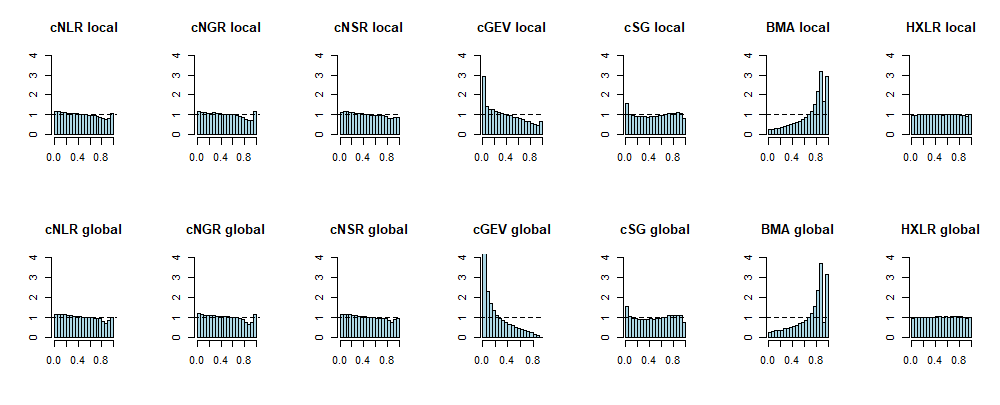} 
\caption{PIT histograms of the basic postprocessing models. \newline}
\label{fig:pit_b}
\end{figure}

\newpage
The PIT histograms of the models are shown in Figure~\ref{fig:pit_b}. Note that a randomized version of the PIT values has to be used for the censored nonhomogeneous regression models. This is due to the fact that the CDF of a censored distribution has a discrete component. The regression models fitted with the R-package \textit{crch} provide the flattest histograms (cNLR, cNGR, cNSR and HXLR). The censored Logistic, Gaussian and Student distribution seem to fit distinctly better than the censored GEV (wet bias) and slightly better than the censored Shifted Gamma distribution. The PIT histogram of the BMA model indicates a dry bias in both versions. \\

\newpage
\section{Extension approaches}
\label{eappr}
This appendix section introduces the alternative extensions for the cNLR model and gives more detail on the approaches presented in the paper. We first present the two basic procedures which have been used to extend the model. Section~\ref{e2} deals with analyses of the topographical and seasonal structure of the ensemble prediction error and the models that result from it. In Section~\ref{e3}, some more basic extension approaches are presented and in Section~\ref{e4}, all the approaches are compared. \\

The aim here is to extend the cNLR model by including topographical, seasonal and prediction situation specific information. The term \textit{prediction situation} describes the by the ensemble given forecasting prerequisites (as for example a classification into wet and dry forecasts based on the ensemble mean and the ensemble standard deviation). In contrast, we use the term \textit{prediction setting} when referring to all three conditions (topography, season and prediction situation) under which a forecast was created. Two extension procedures are pursued: 

\paragraph{Location estimation} 
The first approach integrates additional predictor variables in the moment estimation of the Logistic distribution. The location does not longer depend only on the control member $x_1$ and the ensemble mean $\overline{x}$, but also on additional variables describing the topography, the season and the prediction situation of the prediction setting. A location estimator of the following form emerges:
\begin{align}
\label{loc_esti}
\mu = \beta_0 + \beta_1x_1 + \beta_2\overline{x} + \sum_{j=1}^{d} \zeta_j p_j,
\end{align}
where $p_j$ are interactions and main effects of the additional variables. The $\zeta_j$'s are coefficients and $d$ is the number of additional predictors in use. \\

\paragraph{Weights} 
In the second approach, the regression coefficients are fitted by minimizing a weighted version of the mean CRPS which has the following form: 
\begin{align}
\label{wCRPS}
c(\psi; q) = \sum_{i=1}^{n} w_i(q) CRPS(F_i, y_i;\psi),
\end{align}
where $CRPS(F_i,y_i;\psi)$ refers to the CRPS value of data pair $(F_i, y_i)$ where the predictive distribution $F_i$ depends on the coefficient vector $\psi$. The weight $w_i(q)$ of training data pair $(F_i,y_i)$ depends on the similarity between the conditions in which this forecast-observation pair originated and the prediction setting $q$. Similarities in topography, season and prediction situation have been considered. \citet{similarity1} use such an approach as well, but they have limited themselves to the selection of similar prediction locations calling it a \textit{semi-local} model. In this study, a time weighting of the training data is added. This is done by attaching weight to data from training days which are similar in seasonality and prediction situation. \\

The idea behind the weights called \textbf{NN-weights} is the following: The model is trained with a certain number $W$ of the data pairs which originated under conditions most similar to the prediction setting (the nearest neighbours, NN). These data pairs are weighted equally among each other. The similarity of the conditions of origin is measured with a distance $d$, which depends on the topography, the season or the prediction situation. Let $D_{d}^1(q) \leq D_{d}^2(q) \leq ... \leq D_{d}^n(q)$ be the ordered distances $d(q,q_i)$ between the actual prediction setting $q$ and the $n$ conditions $q_i$, under which the training data pairs originated. Then, the NN-weights with respect to the distance $d$ are defined as:
	
	\begin{equation}
	\label{NN-weights}
	w_i^{NN;d;W}(q) =
	\begin{cases}
	1 & \text{for } d(q,q_i) \leq D_{d}^W(q), \\
	0 & \text{otherwise}.
	\end{cases}
	\end{equation}
	
These weights are a function of the prediction setting $q$ and the conditions of origin of the whole training data set, as the remaining conditions determine, if a setting $q_i$ is under the $W$ nearest neighbours. If several training data pairs originated under conditions with the same distances $d(q,q_i)$, then all the corresponding data pairs are used. \\

The extensions of the cNLR model are compared with the local and the global fit of this very model, the assessment is based on Dataset~1. The target of this study is to develop an area-covering postprocessing method. Therefore, the extended models are trained with data where the predicted month and the predicted station are left out. This simulates the situation where postprocessing must be done outside a weather station, i.e.\ without past local measurements. The training period is the last year ($12$ months) before the prediction month. Consequently, the model performance can only be assessed with the data from $2017$ and $2018$.\\

\subsection{Structure ensemble prediction error}
\label{e2}
This section presents dependencies of the ensemble prediction error on the topography, season and prediction situation. For this purpose, the precipitation amount $(Y_{s,t})_{s \in D, t \in T}$ at location $s$ on day $t$ is expressed as a function $f(\boldsymbol{x}_{s,t},s,t)$ of the location, the day and the corresponding ensemble forecast~$\boldsymbol{x}_{s,t}$. $D \subseteq \mathbb{R}^2$ is a set with the coordinates of Switzerland and $T$ is a set of dates. To find a suitable function $f(\boldsymbol{x}_{s,t},s,t)$, the ensemble prediction error is described by approximating the bias of the ensemble mean as a function $\tau(\boldsymbol{x}_{s,t},s,t):\mathbb{R} \times D \times T \rightarrow \mathbb{R}$, such that
\begin{equation}
f(\boldsymbol{x}_{s,t},s,t) = \overline{x_{s,t}} + \tau(\boldsymbol{x}_{s,t},s,t). 
\end{equation} 

\subsubsection{Topographical structure}
\label{TOPOstr}
\begin{figure}[h]
	\centering
	\includegraphics[width=0.41\textwidth]{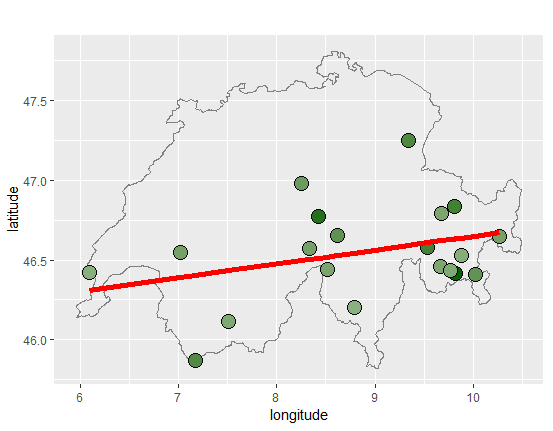}
	\includegraphics[width=0.4\textwidth]{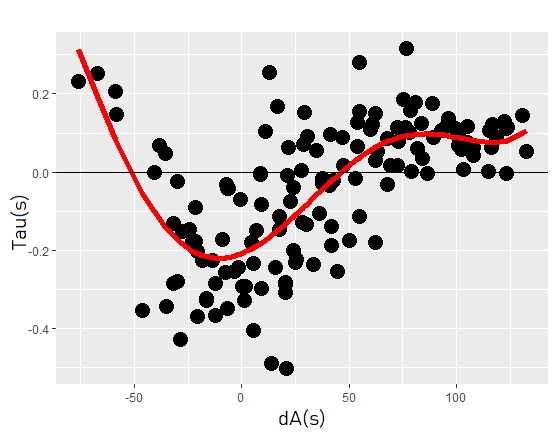}
	\caption{\textbf{Left:} Estimator $\hat{a}$ of the "Alp-line", \quad \textbf{Right:} Scatter-plot of the station-wise means of $y_{s,t}~-~\overline{x_{s,t}}$ versus the distance to the Alps of these same stations.}
	\vspace{0.5cm}
	\label{fig:Tau1}
\end{figure}

To get an idea of the topographical structure of the ensemble mean bias, the station-wise means of $y_{s,t} - \overline{x_{s,t}}$ are illustrated on a map (Figure~3 in the main paper, we use the data from $2016$ out of Dataset~1). It seems reasonable to assume that the bias of the ensemble mean depends on the distance between the prediction location and the Alps. For this reason, an "Alp-line" is defined such that it is possible to quantify the distance between a prediction site and the Alps. To define a line, the latitude of the $20$ highest stations of Dataset~1 is modelled as a linear function of the longitude. A linear regression illustrated in Figure~\ref{fig:Tau1} (left) provides the coefficients such that the following estimator of the "Alp-line" $a$ emerges ($l$ is the longitude):
\begin{equation}
\hat{a}: \mathbb{R} \rightarrow \mathbb{R}, \quad l \mapsto 45.78010 + 0.08678 \cdot l.
\end{equation}

The signed distance to the Alps of location $s$ is then expressed through: 
\begin{equation}
d_A(s) = 
\begin{cases}
||s-p_A(s)|| & \text{for } \text{s north of } \hat{a}, \\
-||s-p_A(s)|| & \text{for } \text{s south of } \hat{a},
\end{cases}
\end{equation}
with $p_A(s)$ the orthogonal projection of $s$ onto the "Alp-line" $\hat{a}$. Figure~\ref{fig:Tau1} (right) depicts a scatter-plot of the station-wise means of $y_{s,t} - \overline{x_{s,t}}$ versus the distance to the Alps of these same stations. To find a function which describes the dependency between those two, the performances of different polynomial fits are compared with F-Tests. There is a significant improvement until the fifth degree, such that the topographical dependency of $\tau$ is expressed as
\begin{equation}
\tau(s) = \sum_{i=0}^{5}\beta_id_A(s)^i.
\end{equation} 
The fitted $\tau(s)$ is illustrated with the red line in Figure~\ref{fig:Tau1} (right). \\

An approach using the distance to the Alpline in the location estimation depends on a lot of coefficients: The number of highest stations used in the fitting of the "Alp-line" has to be determined and the "Alp-line" itself depends on two coefficients. The final location estimation needs seven coefficients more. For this reason, the alternative approach based on the DEM variable has been introduced, the motivating steps are shown in the main paper. \\

On the one hand, the variable DEM in the resolutions $31km$ and $15km$ were added as predictors $p_j$ in the location estimation of the censored Logistic distribution, resulting in the following four models:
\begin{align}
\text{\textbf{DEMm15:}} \quad \mu = &\beta_0 + \beta_1x_1 + \beta_2\overline{x} + \zeta_1 DEM_{15km}, \\
\text{\textbf{DEMm31:}} \quad \mu = &\beta_0 + \beta_1x_1 + \beta_2\overline{x} + \zeta_1 DEM_{31km}, \\
\text{\textbf{DEMmlin:}} \quad \mu = &\beta_0 + \beta_1x_1 + \beta_2\overline{x} + \zeta_1 DEM_{31km} + \zeta_2 DEM_{15km}, \\
\text{\textbf{DEMmint:}} \quad \mu = &\beta_0 + \beta_1x_1 + \beta_2\overline{x} + \zeta_1 DEM_{31km} + \zeta_2 DEM_{15km} \\ &+ \zeta_3 DEM_{15km}:DEM_{31km}. 
\end{align}
Furthermore, the two variables were used to generate weighted versions of the cNLR model. For this purpose, different distances with respect to the $DEM$ variables(s) are defined. They rate the similarity between the location $s_j$ of station $j$ out of the training data set and the prediction location $s$:
	\begin{align*}
	d_{DEM15}(s, s_j) &= | DEM_{15km}(s) - DEM_{15km}(s_j) |, \\
	d_{DEM31}(s, s_j) &= | DEM_{31km}(s) - DEM_{31km}(s_j) |.
	\end{align*}
The distance $d_{DEM31}$ is the distance introduced in the main paper. A third distance is based on both variables. To ensure that they are weighted equally, ranks are used. The station with the $r$-smallest distance $d_{DEM15}(s,s_j)=D_{d_{DEM15}}^{(r)}(s)$ gets rank $R^{d_{DEM15}}(s,s_j)=r$. For congruent distances, middle ranks are used. The same ranking is done for $d_{DEM31}(s,s_j)$. The resulting two ranks are summed and used to define the following distance: 	
	\begin{align*}
	d_{DEM}(s, s_j) &= R^{d_{DEM15}}(s,s_j) + R^{d_{DEM31}}(s,s_j).
	\end{align*} 

The three distances are used to generate weighted extensions of the cNLR model: $d_{DEM15}$ is used in the model \textbf{DEMw15}, $d_{DEM31}$ in the model \textbf{DEMw31} and $d_{DEM}$ in the model \textbf{DEMw(both)}. We fit the model with the training data originating from the $L$ most similar stations and use $d_{DEM}$ to show exemplary how the weights in (\ref{wCRPS}) are defined within this approach. Let $D_{d_{DEM}}^{(1)}(s) \leq D_{d_{DEM}}^{(2)}(s) \leq ... \leq D_{d_{DEM}}^{(m)}(s)$ be the ordered distances $d_{DEM}(s,s_j)$ between the $m$ stations from the training data and the prediction location $s$. Let $s_i$ be the location of the station where forecast-observation pair $(F_i, y_i)$ originated. Then, the NN-weights with respect to the distance $d_{DEM}$ are defined as:

\begin{equation}
\label{NNwDEM}
w_i^{NN;d_{DEM};L}(s) =
\begin{cases}
1 & \text{for } d_{DEM}(s,s_i) \leq D_{d_{DEM}}^{(L)}(s), \\
0 & \text{otherwise}.
\end{cases}
\end{equation}
\vspace{0.2cm}

The NN-weights for the other two distances are defined analogously. 

\newpage
\subsubsection{Seasonal and prediction situation dependent structure}
\label{seasonal}
\begin{figure}[h]
	\centering
	\includegraphics[width=0.6\textwidth]{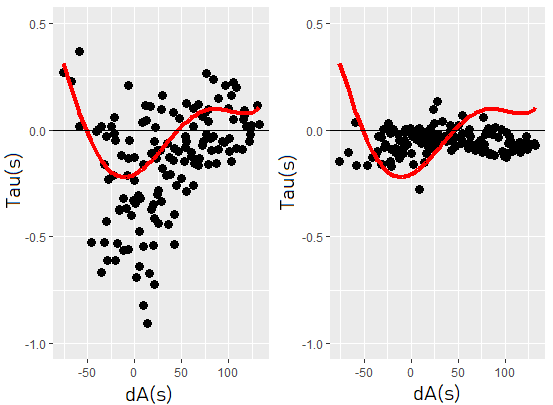}
	\caption{Station-wise means of $y_{s,t} - \overline{x_{s,t}}$ versus the distance to the Alps of these same stations for November 2016 (left) and December 2016 (right).}
	\vspace{0.1cm}
	\label{fig:NovDec}
\end{figure}

To subsequently implement suitable models, some difficulties regarding seasonality and prediction situation are pointed out here. Figure~\ref{fig:NovDec} shows the same plot as Figure~\ref{fig:Tau1} (right), but the station-wise means of $y_{s,t} - \overline{x_{s,t}}$ are calculated monthly instead of over the whole year. The months of November and December $2016$ are shown as examples, the red line is the same fifth order polynomial as in Figure~\ref{fig:Tau1}, fitted with the whole year of data. 

Although November and December $2016$ are neighbouring months, there is a big difference between the scatter-plots. This can be explained by examining how "wet" the two months were: We have $4110$ observations for November $2016$, in December there are $4275$. In November, it did not rain in $54\%$ of the observations, in December in almost twice as many, in $91\%$. To account for the differences between wet and dry periods, ensemble prediction situations are divided into two groups - wet and dry forecasts. The following distinction based on the ensemble mean $\overline{x}$ and the ensemble standard deviation $SD(\boldsymbol{x})$ is used:
\begin{align}
\text{Dry forecast: } &\overline{x} - SD(\boldsymbol{x}) \leq 0.1, \\
\text{Wet forecast: } &\overline{x} - SD(\boldsymbol{x}) > 0.1.
\end{align}
$0.1$ is used instead of $0.0$, such that ensemble predictions with a slightly positive mean and a small standard deviation are defined as dry forecasts as well. \\

\begin{figure}[h]
	\centering
	\includegraphics[width=0.6\textwidth]{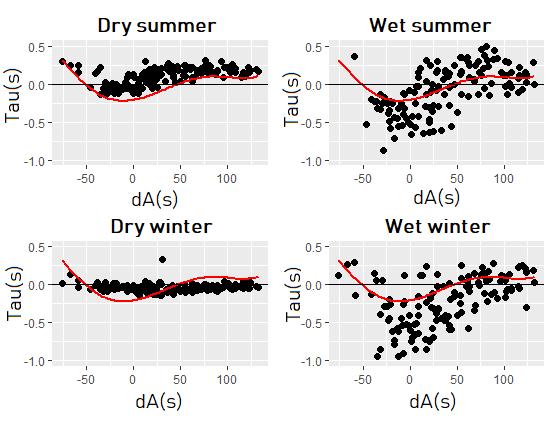}
	\caption{Station-wise means of $y_{s,t} - \overline{x_{s,t}}$ versus the distance to the Alps of these same stations after splitting the data into summer and winter months such as dry and wet forecasts.}
	\vspace{0.2cm}
	\label{fig:Wetdry}
\end{figure}

For Figure~\ref{fig:Wetdry}, the station-wise means of $y_{s,t} - \overline{x_{s,t}}$ are calculated after dividing the data into summer (May-October) versus winter (November-April) months and dry versus wet forecasts. The red line is still the same fifth order polynomial as in Figure~\ref{fig:Tau1}, fitted with the whole year of data. The four graphs illustrate that the dependency of the ensemble mean bias on the distance to the Alps differs for the four scenarios. In dry forecasting situations during winter, the bias is constantly slightly negative and does not seem to depend on the distance to the Alps. Also in dry summer, the mean bias per station is small, but there seems to be some polynomial structure. In wet forecasting situations during summer, the ensemble mean seems to overpredict the precipitation amount in the Alps and to underpredict it in the lowland. For wet forecasting situations during winter, the ensemble mean appears to overpredict the precipitation amount all over the place. \\

To use these findings but avoid sharp distinctions between wet and dry forecasts such as summer and winter months, the following approach to model $\tau(\boldsymbol{x}_{s,t},s,t)$ is proposed:
\begin{equation}
\label{taudef}
\tau(\boldsymbol{x}_{s,t},s,t) = \sum_{i=0}^{5} \beta_i d_A(s)^i
+ \overline{x_{s,t}}*sin(m),
\end{equation}
where $m=\frac{month(t) \cdot \pi}{12}$. The expression $month(t)$ denotes the number of the month day $t$ lies in with $1$~=~January, $2$~=~February,$...$, $12$~=~December. This function enables the dependency of the ensemble mean bias on the interaction and the main effects of the prediction situation and the season. \\

\newpage
In the definition of the $\tau$ function in (\ref{taudef}), a first approach to deal with the seasonal difficulties has been introduced. The seasonal extension approach called \textbf{Sine} is based on this, the location of the censored Logistic distribution is fitted by 

\begin{equation}
\mu = \beta_0 + \beta_1 x_1 + \beta_2 \overline{x} + \zeta_1 sin(m) + \zeta_2 \overline{x}:sin(m),
\end{equation}

where $m$ is the by $\frac{\pi}{12}$ multiplied number of the predicted month and $\overline{x}:sin(m)$ is the interaction between $sin(m)$ and $\overline{x}$. While the Sine model extends the location estimation of the cNLR model, weighted extensions based on the season and the prediction situation have been tested as well. The first weighted approach (called \textbf{Sit}) is another attempt to integrate the observations from Section \ref{seasonal} into the model. For this, the data is split into four parts. The allocation \textit{Sit} of a prediction setting $q$ with ensemble forecast $\boldsymbol{x_q}$ to one of these four parts depends on the season and prediction situation of $q$:
\begin{equation}
Sit(q) =
\begin{cases}
\text{Dry summer} & \text{for $q$ in May-October and } \overline{x_q} - SD(\boldsymbol{x_q}) \leq 0.1, \\
\text{Wet summer} & \text{for $q$ in May-October and } \overline{x_q} - SD(\boldsymbol{x_q}) > 0.1, \\
\text{Dry winter} & \text{for $q$ in November-April and } \overline{x_q} - SD(\boldsymbol{x_q}) \leq 0.1, \\
\text{Wet summer} & \text{for $q$ in November-April and } \overline{x_q} - SD(\boldsymbol{x_q}) > 0.1.
\end{cases}
\end{equation}

The Sit model puts weight only on training data originating from the same \textit{Sit} as the prediction setting. Let $Sit(q_i)$ be the allocation of training data pair $(F_i,y_i)$ and $Sit(q)$ the one of the predicted setting. The following weights are used in (\ref{wCRPS}):
\begin{equation}
w_i(q) =
\begin{cases}
1 & \text{for } Sit(q_i) = Sit(q), \\
0 & \text{otherwise}. 
\end{cases}
\end{equation}

This weighting does not use NN-weights as defined in (\ref{NN-weights}), but the next does. The approach named \textbf{Ensemble characteristics} is inspired by the paper of \citet{similarity1}. Here, the model training is done with data from days which had a similar ensemble prediction situation. This is quantified using the ensemble mean and the ensemble standard deviation. These two ensemble quantities vary temporally and spatially, which is why the training data cannot be combined into groups. The ensemble similarity between the prediction setting $q$ and the conditions $q_i$, under which training data pair $i$ emerged, is measured with the following distance: 
\begin{align*}
d_{enschar}(q,q_i) = \sqrt{(\overline{x_{q}} - \overline{x_{q_i}})^2 + (SD(\boldsymbol{x_{q}}) - SD(\boldsymbol{x_{q_i}}))^2}.
\end{align*}

The weights used in the Ensemble characteristics approach are defined as in (\ref{NN-weights}) with $d(q,q_i)~=~d_{enschar}(q,q_i)$. \\

Finally, the Pretest approach of the main paper is combined with the Sit approach, resulting in \textbf{Pretest and wet/dry situation}. This approach combines the Pretest with the wet/dry part of the Sit approach. It works like the Pretest approach, but first splits the whole data into wet and dry prediction situations. 

\subsection{Basic extensions}
\label{e3}
\label{basic_ext}

The extension approaches from Chapter \ref{e2} are compared with some more basic or general models presented in the following. The first two approaches extend cNLR by adding topographical predictors to the location estimation, the third is a weighted approach based on the euclidean distance. The fourth approach is a more basic way to introduce some seasonality into the models.

\paragraph{Location estimation: Principle Component Regression (PCA)}
\textbf{\\}
When extending the location estimation of the cNLR model by topographical variables, two issues have to be faced: The number of available variables is large and the variables are not independent of each other. One possible solution approach is \textit{Principle Component Regression} (see \citeauthor{PCA} \citeyear{PCA}). The idea is to reduce the matrix of possible predictor variables by summarizing the variables to $p$ \textit{Principle Components} (PC), which are linearly independent of each other. The components correspond to the eigenvectors with the $p$ largest eigenvalues from the spectral expansion 
of the variance-covariance matrix of the predictor variables, each is a linear combination of the predictors. 
In this paper, four Principle Component Regressions are applied to the data, they differ in the number of principle components in use:
\begin{itemize}
	\item PCA95: Use principle components explaining 95$\%$ of the variance between the topographical variables. This needs $p=34$ components.
	\item PCA90: Use principle components explaining 90$\%$ of the variance ($p=23$).
	\item PCA75: Use principle components explaining 75$\%$ of the variance ($p=10$).
	\item PCA50: Use principle components explaining 50$\%$ of the variance ($p=4$).
\end{itemize} 
Scores associated with the $p$ components are added to the covariables for location estimation:
\begin{align*}
\mu = \beta_0 + \beta_1 x_1 + \beta_2 \overline{x} + \zeta_1 pcs_1 + \zeta_2 pcs_2 + ... + \zeta_p pcs_p,
\end{align*}
with $pcs_i$ denoting the score associated with the $i$-th principle component. \\

\paragraph{Location estimation: Add interactions with the \textit{glinternet} algorithm (GLI)} 
\textbf{\\}
Another approach to extend the cNLR model is by adding topographical variables both as additive factors and combined as interactions. To decide which of the variables should be used, the \textit{glinternet} algorithm is applied - a method for learning linear pairwise interactions satisfying strong hierarchy introduced by \citet{glinternet}. \\

The workhorse of \textit{glinternet} is a \textit{group-lasso} regularization, which is a more general version of the \textit{lasso}. The \textit{lasso} regularization was introduced by \citet{glinternet2} for a regression set up with the data pairs $(\boldsymbol{x_i}, y_i)$, where $\boldsymbol{x_i}~=~(x_i^1,x_i^2,...,x_i^d)$ are the regressor variables and $y_i$ the response for observation $i$ with $i=1,2,...,n$. The ordinary least square (OLS) estimates, which are obtained by minimizing the residual squared errors, often lead to estimates with a low bias but large variance. By reducing the number of predictors, the variance can be decreased and a model which is easier to interpret arises. In a \textit{group-lasso} regularization, groups of regressor variables are allowed to be selected into or out of a model together. We assume that there are $g$ groups of predictor variables and denote the feature matrix for group~$k$ as $\boldsymbol{x^k}$. $\boldsymbol{y}$ denotes the vector of observations. The definition of the \textit{group-lasso} estimates $(\hat{\alpha}, \hat{\boldsymbol{\beta}})$  with $\hat{\boldsymbol{\beta}} = (\hat{\beta_1}, \hat{\beta_2}, ..., \hat{\beta_g})^T$ by \citet{glinternet} is:
\begin{align*}
\underset{\alpha, \boldsymbol{\beta}}{\mathrm{argmin}} \frac{1}{2} \left|\left| \boldsymbol{y} - \alpha \boldsymbol{1} - \sum_{k=1}^{g} \boldsymbol{x^k} \beta_k \right|\right|^2_2 + \lambda \sum_{k=1}^{g} \gamma_j||\beta_j||_2,
\end{align*} 
where $||\boldsymbol{x}||_2$ of the vector $\boldsymbol{x}=(x_1,x_2,...,x_n)$ is the $n$-dimensional euclidean norm $\sqrt{\sum\limits_{i=1}^{n} x_i^2}$. The parameter $\lambda$ controls the amount of regularization. Decreasing $\lambda$ reduces the regularization and increases the amount of predictor variables in use. The $\gamma$'s allow to penalize the groups of variables to different extents. \\

The \textit{glinternet} algorithm of \citet{glinternet} fits a \textit{group-lasso} on a candidate set of interactions and their associated main effects. For the regularization parameter $\lambda$, a grid of values is used. The algorithm starts with the regularization parameter $\lambda = \lambda_{max}$ for which all estimates are zero. Then, $\lambda$ is decreased with the result that more variables and their interactions are added to the model. The model is obeying strong hierarchy such that an interaction can only be present if both of its main effects are present as well. \\

Implementation: The first argument of the R-function \textit{glinternet} is a matrix with possible predictor variables, here specified as the topographical variables. The algorithm examines the influence of these variables on a response, which is specified as a second argument of the function. Since we want to investigate the influence of the topography on the prediction error, we select the observed precipitation amount minus the ensemble mean ($y_{s,t} - \overline{\boldsymbol{x}_{s,t}}$) as response variable. The number of variables to find is specified as $20$. Again, only the data from $2016$ out of Dataset~1 is used to find the predictors. The implemented model tries first to use all main effects and interactions proposed by \textit{glinternet} as $p_j$. If it runs into numerical optimization problems, $\lambda$ is increased such that some variables fall out. This procedure is repeated until a model fit is achieved. The $9$ first sets of the output of \textit{glinternet} are used, they are presented in Figure~\ref{fig:glivar}. It stands out, that \textit{glinternet} also attaches the most importance to $DEM_{31km}$ and $DEM_{15km}$, which supports the $DEM$-variable approach introduced in the main paper.  

\begin{figure}[h]
	\centering
	\begin{tabular}{p{1.5cm} p{6cm} p{5cm} r r r }
		Set & Main effects & Interactions \\ 
		\hline 
		$GLI_1$ & $ DEM_{31km}$  &  \\ 
		\hline 
		$GLI_2$ & $ DEM_{31km}, DEM_{15km} $ &  \\ 
		\hline 
		$GLI_3$ & $ DEM_{31km}, DEM_{15km}, TPI_{3km} $ &  \\ 
		\hline 
		$GLI_4$ & $ DEM_{31km}, DEM_{15km}, TPI_{3km}, \newline DEM_{7km} $ &  \\ 
		\hline 
		$GLI_5$ & $ DEM_{31km}, DEM_{15km}, TPI_{3km}, \newline DEM_{7km}, Sx11_{3km} $ &  \\ 
		\hline 
		$GLI_6$ & $ DEM_{31km}, DEM_{15km}, TPI_{3km}, \newline DEM_{7km}, Sx11_{3km} $ & $ Sx7_{250m}*Sx11_{3km} $ \\ 
		\hline 
		$GLI_7$ & $ DEM_{31km}, DEM_{15km}, TPI_{3km},  \newline DEM_{7km}, Sx11_{3km}, Sx12_{25m}, Sx13_{3km} $ & $ Sx7_{250m}*Sx11_{3km} $ \\ 
		\hline 
		$GLI_8$ & $ DEM_{31km}, DEM_{15km}, TPI_{3km}, \newline DEM_{7km}, Sx11_{3km}, Sx13_{3km} $ & $ Sx7_{250m}*Sx11_{3km}, \newline Sx12_{25m}*slope_{25m}, \newline Sx13_{25m}*DEM_{7km} $ \\
		\hline 
		$GLI_{9}$ & $ DEM_{31km}, DEM_{15km}, TPI_{3km}, \newline  DEM_{7km}, Sx13_{3km} $ & $ Sx7_{250m}*Sx11_{3km}, \newline Sx12_{25m}*slope_{25m}, \newline Sx13_{25m}*DEM_{7km}, \newline Sx5_{250m}*Sx11_{7km}, \newline Sx11_{3km}*DEM_{7km}, \newline  slope_{3km}*DEM_{7km} $ \\	
		\hline
	\end{tabular} 
	\caption{Sets of \textit{glinternet} predictor variables.}
	\vspace{0.5cm}
	\label{fig:glivar}
\end{figure} 

\paragraph{Weights: Euclidean distance (3DIM and SPA)} 
\textbf{\\}
This weighted approaches select the training data depending on the euclidean distance between its station of origin and the prediction location. The models are fitted with the training data from the $L$ closest stations (as in the weighted \textit{DEM} models). The spatial euclidean distance between the prediction location $s$ and the location $s_j$ of station $j$ is defined as: 
\begin{align*}
d_{spatial}(s,s_j) &= \sqrt{(lat(s)-lat(s_j))^2 + (long(s)-long(s_j))^2},
\end{align*}
with $lat(s)$ denoting the latitude of location $s$ and $long(s)$ its longitude. If a three dimensional version of the euclidean distance is used, the results are almost similar to the ones with $d_{spatial}$. This is due to the fact that the magnitude of the height is much smaller than those of the latitude and longitude. For this reason, ranks are used. Let us define the following three distances:
\begin{align*}
d_{lat}(s,s_j) &= |lat(s)-lat(s_j)|, \\
d_{long}(s,s_j) &= |long(s)-long(s_j)|, \\
d_{h}(s,s_j) &= |h(s)-h(s_j)|,
\end{align*}
where $h(s)$ is the height of location $s$. \\
The station with the $r$-smallest distance $d_{lat}(s,s_j)=D_{d_{lat}}^r(s)$ gets rank $R^{d_{lat}}(s,s_j)=r$. If some distances are equal, middle ranks are used. The same ranking is done for $d_{long}(s,s_j)$ and $d_{h}(s,s_j)$. The resulting three ranks are summed and lead to the following definition of a distance: 
\begin{align*}
d_{3Dim}(s,s_j) = R^{d_{lat}}(s,s_j) + R^{d_{long}}(s,s_j) + R^{d_{h}}(s,s_j).
\end{align*}

The NN-weights for the SPA and the 3DIM approach are defined as for the weighted \textit{DEM} models (see Definition (\ref{NNwDEM})) using the distances $d_{spatial}(s,s_j)$ and $d_{3DIM}(s,s_j)$. Such an approach based on the euclidean distance is also proposed by \citet{similarity1}. 

\begin{figure}[h]
	\centering
	\includegraphics[width=0.43\textwidth]{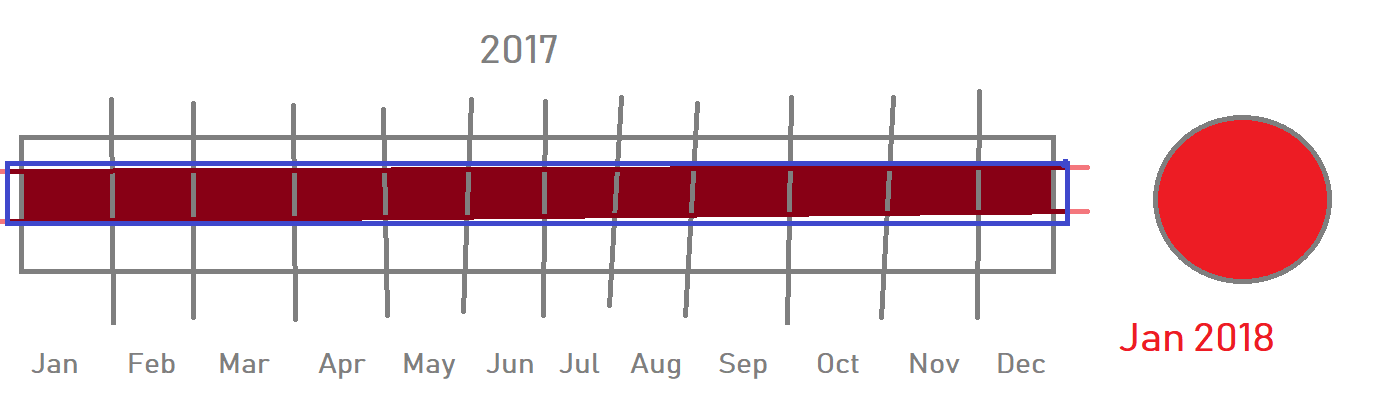}
	\includegraphics[width=0.43\textwidth]{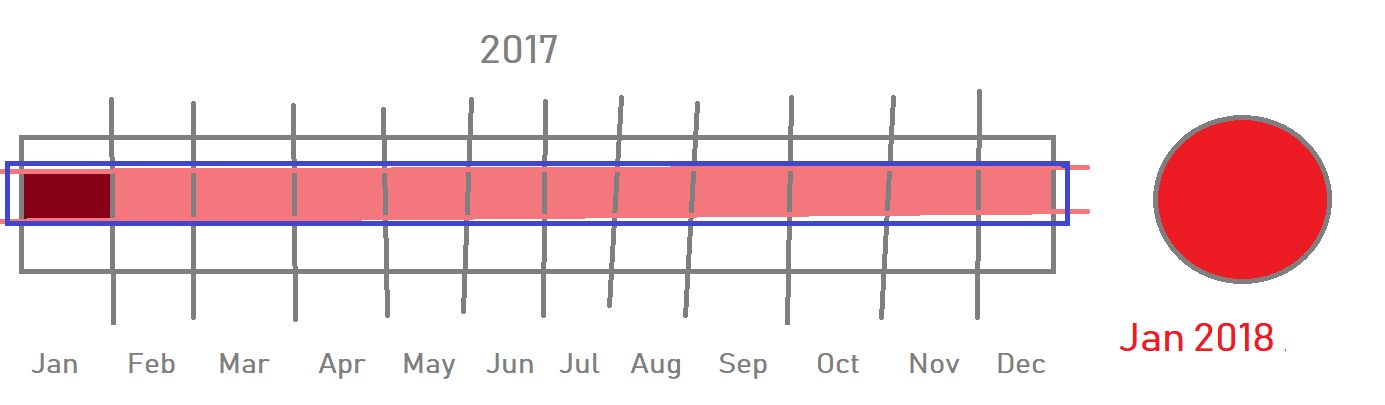}
	\includegraphics[width=0.43\textwidth]{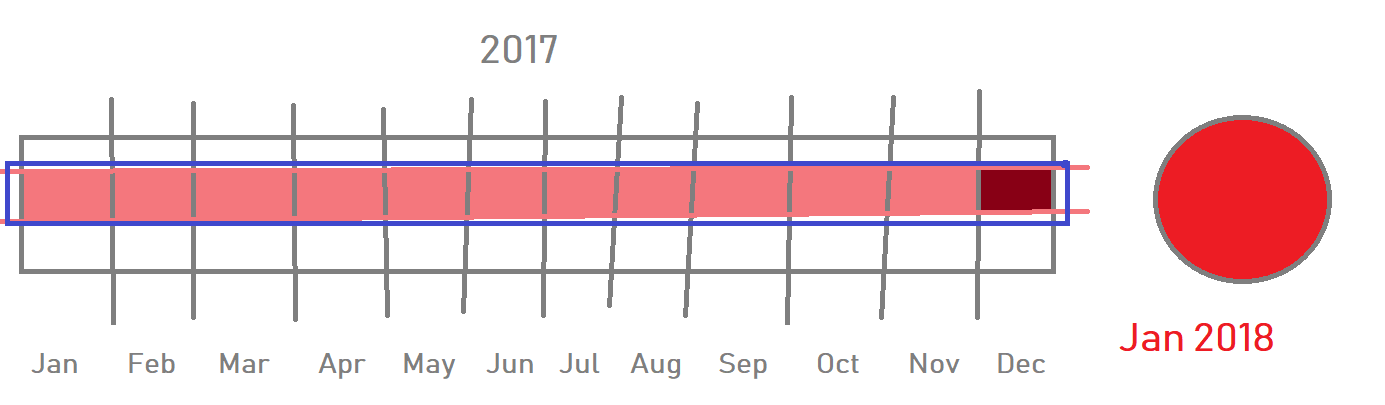}
	\includegraphics[width=0.43\textwidth]{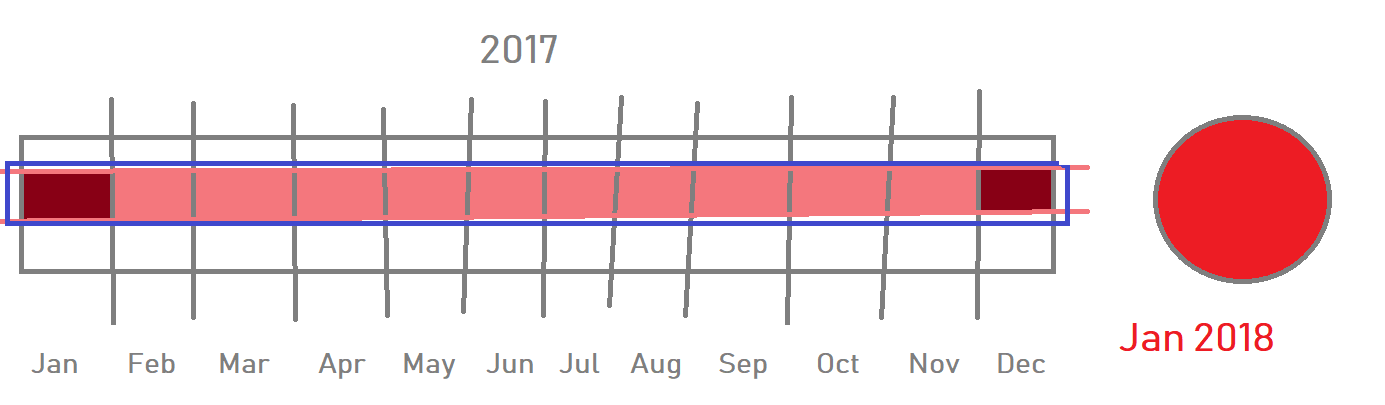}
	\caption{Illustration of the basic seasonal approaches for January $2018$.}
	\vspace{0.5cm}
	\label{fig:Seas_basic}
\end{figure}

\paragraph{Basic seasonality} 
\textbf{\\}
Principally, the extension models are fitted monthly with the previous year of training data. This would refer to the illustration top left in Figure~\ref{fig:Seas_basic}. For comparison, the models are also trained with only one or two months from the same season. These basic seasonality approaches are applied in three versions: The first trains with the month before the prediction month (Figure~\ref{fig:Seas_basic}, bottom left), the second with the same month as the prediction month in the year before (top right) and the third with both of these months (bottom right).

\newpage
\subsection{Results}
\label{e4}
The first part of this chapter compares the performances of the topographical extensions. They are combined with the basic seasonality approaches explained in Figure~\ref{fig:Seas_basic}. Afterwards, the best topographical models are joined with the advanced seasonal approaches. In this paper, we use the square root of the predictions and the observations for the postprocessing. The results of this chapter base on the transformed values such that the magnitude is not the same as in the result section of the main paper, where we reversed the transformation.

\subsubsection{Topographical extensions with basic seasonality}
\label{ext1}
\begin{figure}[h]
	\centering
	\includegraphics[width=1\textwidth]{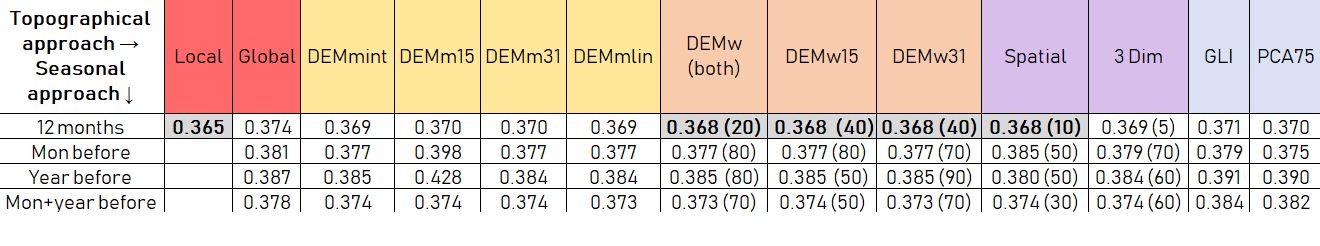}
	\includegraphics[width=0.6\textwidth]{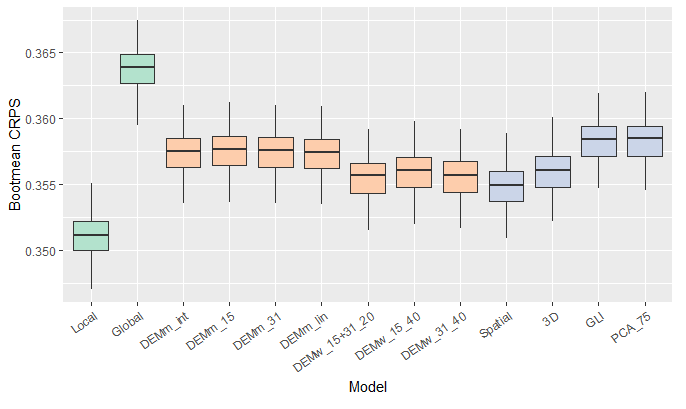}
	\caption{\textbf{Top:} Mean CRPS values of the models combining the topographical and the basic seasonality approaches. For the weighted models, there is an additional number in brackets showing for which $L$ the best performance has been achieved. \textbf{Bottom:} Boxplot of the bootstrapped (250 samples) mean CRPS values of the same topographical models, but only in the version trained with whole previous year.} 
	\label{fig:tab_basicext}
\end{figure}

This section aims to compare the topographical extensions of the cNLR model introduced in Chapter \ref{TOPOstr} and \ref{basic_ext}. The topographical approaches are combined with the training methods of the basic seasonality models. The extensions are fitted for the available months from $2017$ and $2018$ to be assessed afterwards with the test data from the same months (Dataset~1). The table in Figure~\ref{fig:tab_basicext} shows the averages of the CRPS values of this time period. Note that the mean CRPS value of the raw ensemble is $0.390$. 

The local version of the cNLR model has still the lowest mean CRPS value. This model is not further extended as it is not able to perform area-covering postprocessing. Its assessment serves only as a benchmark. For the Principle Component models, only the model using the components explaining 75$\%$ of the variance in the topographical variables is depicted, this version has the lowest mean CRPS value. It is obvious, that the models trained with only one or two months of the same season do (overall) not perform as well as the models that use the data from the whole previous year. For this reason, the boxplot of the bootstrapped mean CPRS values at the bottom of Figure~\ref{fig:tab_basicext} is limited to the models trained with the data from the whole previous year. The boxplot shows that all topographical extensions outperform the global version of the cNLR model but are not able to level with the local version. The lowest mean CRPS values are achieved by DEMw31 and SPA. As a reminder: The DEMw31 model is a weighted model based on the similarity concerning the topographical variable $DEM_{31km}$. The weighting in the SPA model bases on the euclidean distance between the prediction sites. As already explained, we assess here with Dataset~1 and the square root transformation of the precipitation amount. In the later validation with Dataset~2 and after inverting the square root, the DEMw31 model was able to level the performance of the local model. It was possible to show the same for Dataset~1 after inverting the square root transformation.

\newpage
\subsubsection{Topographical extensions with advanced seasonality}
\label{ext2}
This section compares the advanced seasonal extensions of the DEMw31 and the SPA model. Both models are using NN-weights depending on the prediction location. The regression coefficients are fitted with the training data from a given number $L$ of stations, which are most similar to the prediction location. The following list explains how the different combinations of the DEMw31 model and the advanced seasonal approaches are implemented. \\

\textbf{Ensemble characteristics:} For a model which combines the DEMw31 weights with the ensemble characteristic weights, a new distance depending on $d_{DEM31}(s, s_j)$ and $d_{enschar}(q, q_i)$ has to be defined. For this reason, both distances are ranked: 
	\begin{itemize}
		\item[-] The station $j$ of the training data set with the $r$-smallest distance $d_{DEM31}(s, s_j)$ to the prediction location $s$ gets rank $R^{d_{DEM31}}(s,s_j)=r$. If some distances are equal, middle ranks are used.
		\item[-] The training data pair $i$ originating under conditions with the $r$-smallest distance $d_{enschar}(q, q_i)$ to the prediction setting $q$ gets rank $R^{d_{enschar}}(q,q_i)~=~r$. In the case of equal distances, middle ranks are used.
	\end{itemize}
This two ranks are summed to build the following distance:
\begin{align*}
d_{DEM31+enschar}(q, q_i) = R^{d_{DEM31}}(s,s_i) + R^{d_{enschar}}(q,q_i),
\end{align*}
where $s$ is the location of the predicted setting $q$ and $s_i$ the one where the training data pair $i$ originated. This distance is used to generate NN-weights as in (\ref{NN-weights}). The model, which has to be fitted for every day separately, uses training data from the $W=1'000,2'000,...,6'000$ most similar conditions of origin. \\

\textbf{Sit and Pretest wet/dry}: These two models have to be fitted daily as well. First, the training data is reduced to the data pairs from the $L=10,20,...,80$ stations with the smallest distances $d_{DEM31}(s,s_j)$ to the predicted location. Afterwards, the seasonal approaches are applied as described in Chapter \ref{seasonal}. \\

\textbf{Sine and Pretest}: The implementation of these models works analogous but the fit can be done per month.  \\ 

Generally, the advanced seasonal extensions of the SPA model are implemented in the same way where $d_{DEM31}(s,s_j)$ is replaced by $d_{SPA}(s,s_j)$. Since several best results were obtained with the boundary numbers of the parameters $L$ and $W$, the variety of the numbers in use had to be widened.\footnote{ The ensemble characteristics extension is tested with $W=500, 1'000, 2'000,...,6'000$, the other extensions additionally with $L=4,5,...,9$.} \\

\begin{figure}[h]
	\centering
	\includegraphics[width=0.5\textwidth]{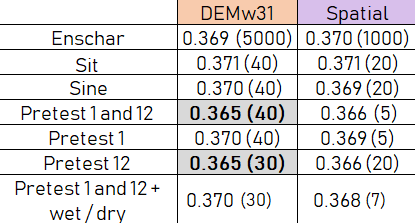}
	\caption{Mean CRPS values of the advanced seasonal extensions of the DEMw31 and the SPA model, based on test data from $2017$ and $2018$. The number in brackets is denoting with which $W$ (Ensemble characteristics) respectively $L$ (other approaches) the models perform best.} 
	\vspace{0.2cm}
	\label{fig:meanCRPS_seas}
\end{figure}

The table in Figure~\ref{fig:meanCRPS_seas} compares the overall performances of the advanced seasonal extensions of the DEMw31 and the SPA model. The best performances are achieved by the DEMw31 model combined with the Pretests 12 (pretesting with the month before, Pretest~2 in main paper) and Pretest 1+12 (pretesting with the month before and the same month of the year before, Pretest~3 in main paper). These models level with the local version of the cNLR model and provide an important outcome: We are able to reproduce the mean CRPS value of the local cNLR model with an area-covering method. Again, we assess here with Dataset~1 and the square root transformation of the precipitation amount. In the later validation with Dataset~2 and after inverting the square root, the DEMw31+Pretest model was able to outperform the local model.  It was possible to show the same for Dataset~1 after inverting the square root transformation.
 
\begin{figure}
	\centering
	\includegraphics[width=1\textwidth]{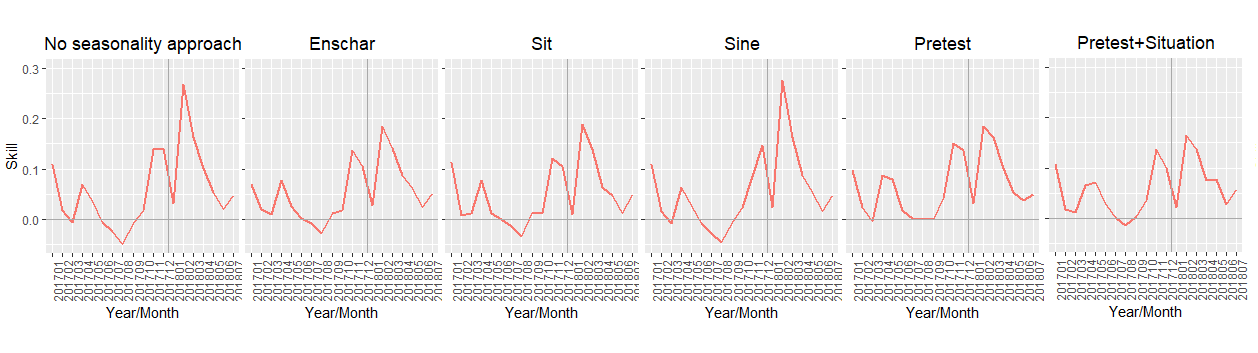}
	\caption{Monthly means of the Skill Score comparing the different seasonal extensions of the DEMw31 model.} 
	\vspace{0.2cm}
	\label{fig:SkillM_DEMw31_seas}
\end{figure}

\newpage
To compare the monthly performances of the advanced seasonal extensions, the monthly Skill Score is depicted in Figure~\ref{fig:SkillM_DEMw31_seas}. Again, the CRPS is used as measure of accuracy and the raw ensemble as reference forecast. The figure is limited to the seasonal extensions of DEMw31. On the left, the DEMw31 model without seasonal extension is given as a baseline. We note that the Pretest model is the only approach without negative skill during summer. In return, the timeline of this approach seems to lose the peaks in other months. 

\end{document}